\documentclass[11pt]{article}
\usepackage{amssymb}
\usepackage{graphics}
\usepackage{epsfig}
\usepackage{a4wide}
\usepackage{cite}
\usepackage{multirow}
\usepackage{color}

\DeclareGraphicsExtensions{.eps}

\newcommand{\cf}[1]{{Fig.\ref{#1}}}

\begin{document}
\title{ Next-to-next-to-leading order QCD corrections to light Higgs pair production via
vector boson fusion in type-II two-Higgs-doublet model }
\author{ Li Wei-Hua, Zhang Ren-You, Ma Wen-Gan, Guo Lei, Ling Liu-Sheng and Li Xiao-Zhou\\
{\small  Department of Modern Physics, University of Science and Technology}  \\
{\small  of China (USTC), Hefei, Anhui 230026, P.R.China}}

\maketitle \vskip 15mm
\begin{abstract}
We present the precision predictions on the pair production of
light, $CP$-even Higgs in weak vector boson fusion (VBF)
up to the QCD next-to-next-to-leading-order (NNLO) at hadron
colliders within the $CP$-conserving type-II two-Higgs-doublet model
(2HDM(II)) by adopting the structure function approach. We
investigate the model parameter dependence, residual uncertainties from the
factorization/renormalization scale, PDFs and $\alpha_s$ on the
integrated cross section at the QCD NNLO, and find that the NNLO QCD
corrections can reduce the scale uncertainty significantly. By
analyzing the kinematic distributions of final Higgs bosons, we can
extract the $CP$-even Higgs resonance via $H^0 \to h^0 h^0$ channel
as a means of probing the extension of the Standard Model (SM) Higgs
sector.
\end{abstract}


\vfill \eject \baselineskip=0.32in

\renewcommand{\theequation}{\arabic{section}.\arabic{equation}}
\renewcommand{\thesection}{\arabic{section}}

\makeatletter      
\@addtoreset{equation}{section}
\makeatother       

\par
\vskip 5mm
\section{Introduction}
\label{sec:intro}
\par
Both ATLAS and CMS collaborations at the Large Hadron Collider (LHC)
have discovered a $126~{\rm GeV}$ neutral boson whose properties are
compatible with the Standard Model (SM) Higgs boson
\cite{ATLAS-2012,CMS-2012,CMS-2013}. The nature of this particle,
including its $CP$ properties and couplings, is currently being
established \cite{CMS-2013,ATLAS-2013}. So the next important step
most probably is the quest for the origin of electroweak symmetry
breaking (EWSB). To achieve this goal, measurement of the Higgs
self-interactions is necessary, which is the only way to reconstruct the
Higgs potential, and determine whether the new particle is the SM
Higgs boson or one of an enlarged Higgs sector of new physics.
Thus, it is useful to explore the implication of the current Higgs
search results on models beyond the SM.

\par
One of the simplest extensions of the SM Higgs sector is the
two-Higgs-doublet model (2HDM) \cite{higgs-guide,thdm-review}. It
predicts the existence of two neutral $CP$-even Higgs bosons ($h^0$
and $H^0$), one neutral $CP$-odd Higgs ($A^0$), and two charge Higgs
bosons ($H^{\pm}$). In addition to their masses, two additional
parameters are introduced in the theory: the ratio of the vacuum
expectation values (VEVs) of the two Higgs doublets $\tan\beta$, and
the mixing angle between the two $CP$-even Higgs fields $\alpha$.
There are many types of 2HDMs, each differing in the way that the
two Higgs doublets couple to the fermions (for a comprehensive
review, see \cite{thdm-review}). In this paper, we only consider the
2HDM of type-II (2HDM(II)), which is designed to avoid
flavor-changing couplings of the neutral Higgs bosons by one Higgs
doublet coupling solely to up-type and the other to down-type
fermions. And this model shares many of the features of the Higgs
sector of the Minimal Supersymmetric Standard Model (MSSM).

\par
To understand the Higgs self-interactions, the only accessible
process is double Higgs production. At the LHC, the most important
SM Higgs boson pair production channels have been systematically
surveyed in Refs.\cite{spira:1212}. The main processes are: (1)
gluon-gluon fusion, $gg \to h^0 h^0$, through heavy-quark loop, (2)
vector boson fusion (VBF) $qq^{\prime} \to V^{\ast} V^{\ast} \to
q''q''' h^0 h^0$, where vector bosons $W/Z$ are radiated off quarks
and fusion to Higgs pair, (3) top-quark pair associated Higgs boson
pair production $q\overline{q}/gg \rightarrow t\bar{t} h^0 h^0$, and
(4) double Higgs strahlung $qq^{\prime} \to h^0 h^0 V$, where Higgs
bosons are radiated off gauge bosons. In the SM, Higgs pair
production via VBF has the second largest cross section and offers a
clean experimental signature of two centrally produced Higgs bosons
with two hard jets in the forward/backward rapidity region. Hence,
it is meaningful to investigate the properties of trilinear Higgs
self-interactions in this clean reaction. In this paper, we focus on
the light $CP$-even Higgs pair production via VBF process $pp \to
V^{\ast} V^{\ast} + 2~jets \to h^0 h^0 + 2~jets$ within the 2HDM(II)
to survey the properties of the trilinear Higgs self-couplings
$\lambda_{h^0 h^0 h^0}$ and $\lambda_{H^0 h^0 h^0}$ appearing in the
Higgs potential \cite{Moretti:2004wa,spira:1212}. In the previous
research works, the VBF Higgs boson pair production process $pp \to
h^0 h^0 + 2~jets$ was surveyed in the 2HDM(II) at the QCD NLO
\cite{Moretti:2007jhep,Figy:nlo}.

\par
Due to the smallness of the QCD interference between the two
inclusive final proton remnants, the VBF single/pair Higgs
production at the leading order (LO) can be viewed as a double
deep-inelastic scattering (DIS) process in a very good
approximation, and the production rate can be computed by adopting
the well-known structure function (SF) approach. Apart from the
interference effect, the SF approach can still be exactly employed
at the QCD next-to-leading order (NLO)
\cite{han:vbfnlo,spira:vbfnlo,6-triple}. Recently, the SF approach
was used to calculate the VBF single/pair Higgs production at hadron
colliders in the SM up to the QCD next-to-next-to-leading order
(NNLO) \cite{big-vbf:prl,big-vbf:prd,vbf-NNLO}. In this paper, we will implement the SF
approach to calculate the VBF Higgs pair production in the 2HDM(II)
up to the QCD NNLO, and provide not only the total cross section,
but also some kinematic distributions of the final Higgs bosons.

\par
The paper is organized as follows. In Sec.\ref{sec:theTHDM}, we give
a brief introduction to the 2HDM(II). The description of the SF
approach and the strategy of the QCD NNLO calculation are presented
in Sec.\ref{sec:CalculationMethod}. In
Sec.\ref{sec:NumericalResults}, we give the numerical results and
focus on the theoretical uncertainty and some kinematic
distributions. A short summary is given in Sec.\ref{sec:summary}.
Finally, we present the analytic expressions for the phase space
element and matrix elements of the VBF Higgs pair
production processes in Appendix.

\vskip 5mm
\section{Two-Higgs-Doublet Model of Type-II }
\label{sec:theTHDM}
\par
The 2HDM contains two scalar $SU(2)_L$ doublets, $\Phi_{1}$ and
$\Phi_{2}$, with weak hypercharge $Y=1$. The most general Higgs
potential with $SU(2)_L \times U(1)_Y$, $Z_2$
and $CP$ symmetries has the form as \cite{higgs-guide}
\begin{eqnarray}
\label{eq:potential} V(\Phi_1,\Phi_2) &=& m^2_{11} \Phi^\dagger_1
\Phi_1 + m^2_{22} \Phi^\dagger_2 \Phi_2 + \frac{1}{2}\lambda_1 (
\Phi^\dagger_1 \Phi_1 )^2 +
\frac{1}{2}\lambda_2 ( \Phi^\dagger_2 \Phi_2 )^2 \nonumber \\
&& + \lambda_3 ( \Phi^\dagger_1 \Phi_1 ) ( \Phi^\dagger_2 \Phi_2 ) +
\lambda_4 ( \Phi^\dagger_1 \Phi_2 ) ( \Phi^\dagger_2 \Phi_1 ) +
\frac{1}{2} \lambda_5 \left[ ( \Phi^\dagger_1 \Phi_2 )^2 + {\rm
h.c.} \right],
\end{eqnarray}
where $m_{11}^2$, $m_{22}^2$ and $\lambda_i~ (i = 1,...,5)$ are all
real parameters, and $\Phi_{1,2}$ transform under the $Z_2$ discrete
symmetry as
\begin{eqnarray}
\Phi_1 \rightarrow \Phi_1,~~~~~~~~\Phi_2 \rightarrow -\Phi_2.
\end{eqnarray}
After EWSB, the neutral components of $\Phi_1$ and $\Phi_2$ acquire
VEVs $v_1/\sqrt{2}$  and $v_2/\sqrt{2}$, respectively, which are determined
by the vacuum conditions of
\begin{eqnarray}
\label{eq:vacuum_stationary}
&& m^2_{11} v_1 + \frac{1}{2}\lambda_1 v^3_1 + \frac{1}{2} \left( \lambda_3
+ \lambda_4 + \lambda_5 \right) v_1 v^2_2 = 0, \nonumber \\
&& m^2_{22} v_2 + \frac{1}{2}\lambda_2 v^3_2 + \frac{1}{2} \left( \lambda_3
+ \lambda_4 + \lambda_5 \right) v_2 v^2_1 = 0,~~~~~~~~~~~~~~~~~~~
\end{eqnarray}
and satisfy $\sqrt{v_1^2 + v_2^2} \equiv v \simeq 246~ {\rm GeV}$
\footnote{In this paper, the VEV $v$ is fixed by the masses of weak gauge
bosons $M_W$, $M_Z$ and Fermi constant $G_F$.}. We parameterize the two
Higgs doublets as
\begin{eqnarray}
\Phi_i =
\left(
\begin{array}{c}
\phi^+_i \\
\frac{1}{\sqrt{2}}(v_i + R_i + i I_i)
\end{array}
\right),~~~~~~~(i=1, 2).
\end{eqnarray}
The Higgs mass matrices are diagonalized by performing the following
rotation transformations:
\begin{eqnarray}
\left(
\begin{array}{c}
H^0 \\
h^0
\end{array}
\right)
= R(\alpha)
\left(
\begin{array}{c}
R_1 \\
R_2
\end{array}
\right),~~~~
\left(
\begin{array}{c}
G^0 \\
A^0
\end{array}
\right)
= R(\beta)
\left(
\begin{array}{c}
I_1 \\
I_2
\end{array}
\right),~~~~
\left(
\begin{array}{c}
G^+ \\
H^+
\end{array}
\right)
= R(\beta)
\left(
\begin{array}{c}
\phi_1^+ \\
\phi_2^+
\end{array}
\right),
\end{eqnarray}
where the rotation matrix $R$ is defined as
\begin{eqnarray}
R(\theta) =
\left(
\begin{array}{cc}
\cos\theta & \sin\theta \\
-\sin\theta & \cos\theta
\end{array}
\right),
\end{eqnarray}
$\alpha$ is the mixing angle between the two $CP$-even Higgs fields
$R_1$ and $R_2$, and $\tan\beta = v_2/v_1$. The fields $G^0$ and
$G^{\pm}$ are Nambu-Goldstone bosons and their three degrees of
freedom are got ``eaten" by the longitudinal components of $Z$ and
$W^{\pm}$ bosons, and induce the masses of weak gauge bosons.
Therefore, the 2HDM predicts five scalar particles: $h^0$, $H^0$,
$A^0$ and $H^{\pm}$. We may choose the following seven independent
``physical" parameters as the inputs of the Higgs sector:
\begin{eqnarray}
m_{h^0},~~ m_{H^0},~~ m_{A^0},~~ m_{H^\pm},~~ \sin\alpha,~~ \tan\beta,~~ v.
\end{eqnarray}
Then the quartic couplings $\lambda_{1,...,5}$ can be expressed in
terms of these physical parameters as
\begin{eqnarray}
\label{Lambda-Mass}
\lambda_1
=
\frac{1}{v^2 \cos^2\beta}
\left(
m_{h^0}^2 \sin^2\alpha  + m_{H^0}^2 \cos^2\alpha
\right),~~~~
\lambda_2
=
\frac{1}{v^2 \sin^2\beta}
\left(
m_{h^0}^2 \cos^2\alpha  + m_{H^0}^2 \sin^2\alpha
\right), \nonumber \\
\lambda_3
=
2 \frac{m_{H^\pm}^2}{v^2} + \frac{\sin 2\alpha}{v^2 \sin 2\beta}
\left( m_{H^0}^2 - m_{h^0}^2 \right),~~~~
\lambda_4
=
\frac{1}{v^2} \left( m_{A^0}^2 - 2 m_{H^\pm}^2 \right),~~~~
\lambda_5
=
-\frac{1}{v^2}m^2_{A^0}.~~
\end{eqnarray}

\par
The tree-level couplings of $h^0$, $H^0$ and $A^0$ to the SM gauge
bosons and fermions with respect to the corresponding couplings of
the SM Higgs boson are presented in Table \ref{tab:2HDMcoupling}. It
should be mentioned that the couplings of the $CP$-even Higgs bosons
$h^0$ and $H^0$ have the same structures as those of the SM Higgs
boson, while the Feynman rules for the $A^0-f-\bar{f}$ interactions
contain an additional factor $i \gamma^5$ since $A^0$ is a
pseudoscalar. We can see from the table that when $(\beta - \alpha)
\to \frac{\pi}{2}$, the couplings of the light $CP$-even Higgs $h^0$
to gauge bosons and fermions approach the corresponding SM ones, and
the couplings of the heavy $CP$-even Higgs $H^0$ to weak gauge
bosons approach zero. Therefore, the $CP$-even Higgs $H^0$ decouples
from the VBF process $pp \to V^{\ast} V^{\ast} + 2~jets
\to h^0 h^0 + 2~jets$ in the SM limit of $(\beta - \alpha) =
\frac{\pi}{2}$. In this work we use $\sin(\beta-\alpha)$ as an input
parameter of the Higgs sector instead of $\sin\alpha$ to manifest
the effects on the Higgs couplings to gauge bosons involved in the
VBF $h^0$ pair production, considering the fact that the Higgs couplings to
weak gauge bosons are proportional to $\sin(\beta-\alpha)$ and
$\cos(\beta-\alpha)$.
\begin{table}
  \centering
  \begin{tabular}{c|c|c|c}
    \hline\hline
     ~~~~~ & ~~~~$WW$, $ZZ$~~~~ & ~~~~up-type quarks~~~~ & down-type quarks, leptons \\
    \hline
    $h^0$ & $\sin(\beta-\alpha)$ & $\cos\alpha/\sin\beta$ & $-\sin\alpha/\cos\beta$ \\
    $H^0$ & $\cos(\beta-\alpha)$ & $\sin\alpha/\sin\beta$ &  $\cos\alpha/\cos\beta$ \\
    $A^0$ &                  $0$ & $i\gamma^5\cot\beta$   &  $i\gamma^5\tan\beta$ \\
    \hline\hline
  \end{tabular}
  \caption{
    Tree-level couplings of the neutral Higgs bosons of the 2HDM(II) to gauge bosons
    and fermions. Each coupling is normalized to the corresponding coupling of
    the SM Higgs boson.}
\label{tab:2HDMcoupling}
\end{table}

\vskip 5mm
\section{Calculation Strategy}
\label{sec:CalculationMethod}
\par
The SF approach is a very good approximation for studing the VBF
processes at hadron colliders, which is accurate at a precision
level well above the typical residual scale and parton distribution
function (PDF) uncertainties \cite{big-vbf:prl}. This approximation
is based on the absence or smallness of the QCD interference between
the two inclusive final proton remnants. The Higgs boson pair
production via VBF is a pure electroweak process at the LO, see
\cf{fig:vbf}. There are two types of topological Feynman diagrams
($t$- and $u$-channel) contributing to the VBF Higgs pair production
at parton level. The cross section is approximately contributed only
by the squared $t$- and $u$-channel amplitudes, while their
interference contribution is below $0.01\%$. Therefore, the VBF
Higgs pair production can be viewed as the double deep-inelastic
scattering (DIS) of two (anti)quarks with two virtual weak vector
bosons independently emitted from the hadronic initial states fusing
into a Higgs boson pair \cite{spira:1212}. The cross section can be
calculated in terms of the charged-current and neutral-current
hadronic structure functions $F_i^{V}(x, Q^2)~ (i = 1, 2, 3,~ V = Z,
W^{\pm})$ by adopting the SF approach \cite{han:vbfnlo}. This method
has been implemented to calculate the NNLO QCD corrections to the
single Higgs production via VBF \cite{big-vbf:prl,big-vbf:prd}.
Analogous to the case of the VBF single Higgs production, the
nonfactorization contribution to the VBF Higgs pair production can
also be safely neglected \cite{big-vbf:prd}. In this paper we adopt
the SF approach to calculate the total inclusive cross section and
differential distributions in the 2HDM(II) at the QCD NNLO for the
VBF Higgs pair production $pp \to V^{\ast} V^{\ast} + 2~jets \to h^0
h^0 + 2~jets$.
\begin{figure}[ht!]
\centering
\includegraphics[scale=1]{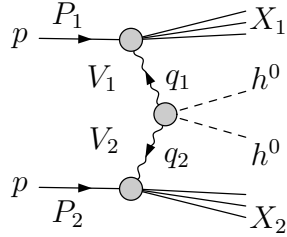}
\caption{Light $CP$-even Higgs pair production via VBF.}
\label{fig:vbf}
\end{figure}

\par
The differential cross section for the VBF Higgs pair production can be
expressed as \cite{big-vbf:prd}
\begin{eqnarray}
\label{eq:sfapproach}
d\sigma
&=&
\sum_{(V_1 V_2)}
\frac{1}{2 S} 2 G_F^2 M^2_{V_1} M^2_{V_2}
\frac{1}{\left( Q^2_1 + M^2_{V_1} \right)^2}
\frac{1}{\left( Q^2_2 + M^2_{V_2} \right)^2}
W_{\mu \nu}^{V_1}(x_1, Q^2_1)
{\cal M}^{\mu \rho}_{V_1 V_2}
{\cal M}^{\ast \nu \sigma}_{V_1 V_2}
W_{\rho \sigma}^{V_2}(x_2, Q^2_2) \nonumber \\
&&
\times
\frac{d^3 \vec{P}_{X_1}}{\left( 2 \pi \right)^3 2 E_{X_1}}
\frac{d^3 \vec{P}_{X_2}}{\left( 2 \pi \right)^3 2 E_{X_2}}
d s_1 d s_2
d PS_2(k_1, k_2)
\left( 2 \pi \right)^4 \delta^4 \left( P_1 + P_2 - P_{X_1} - P_{X_2} - \sum_{j=1,2} k_j \right),~~~~~~
\end{eqnarray}
where $(V_1 V_2) = (ZZ), (W^+W^-), (W^-W^+)$, $G_F$ is the Fermi
constant, $\sqrt{S}$ is the center-of-mass energy of the hadron
collider, $d PS_2(k_1, k_2)$ represents the phase space of the final
two Higgs bosons, ${\cal M}^{\mu \nu}_{V_1 V_2}$ stands for the
matrix element for the VBF subprocess $V_1(-q_1) + V_2(-q_2)
\rightarrow h^0(k_1) + h^0(k_2)$, the physical scale $Q$ is given by
$Q^2_i = -q^2_i$ for $x=x_i~(i=1,2)$ and $x_i = Q^2_i/(2P_i \cdot
q_i)$ are the usual DIS variables, and $s_i = (P_i + q_i)^2$ are the
invariant mass of the $i$-th proton remnant. At the end of
Eq.(\ref{eq:sfapproach}) there includes the four-body final state
phase space element for the VBF Higgs pair production process, which
is expressed explicitly in Appendix \ref{sec:PhaseSpace}.

\par
\begin{figure}[ht!]
\centering
\includegraphics[scale=0.6]{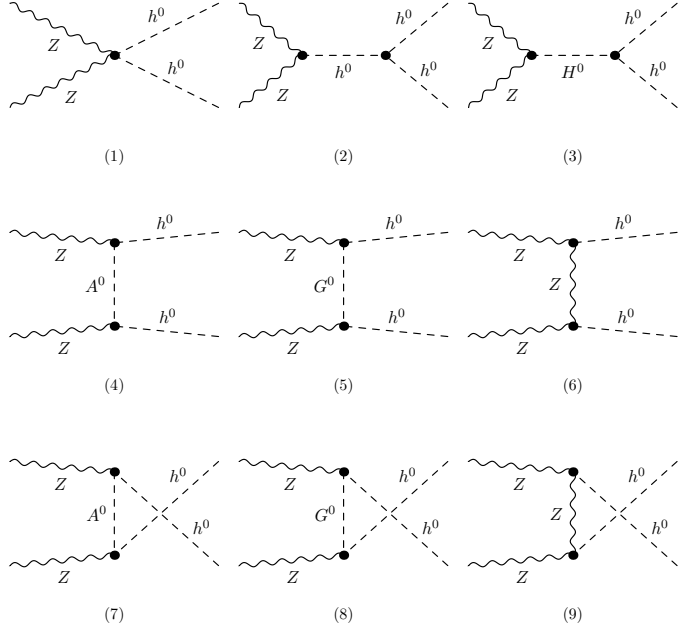}
\caption{Feynman diagrams for the $ZZ \rightarrow h^0 h^0$ process.}
\label{fig:zzfusion}
\end{figure}

The DIS hadronic tensor $W_{\mu \nu}^V(x, Q^2)$ can be expressed in
terms of the standard DIS structure functions $F_j^V(x_i, Q_i^2)~
(i=1,2,~j = 1, 2, 3)$ as
\begin{eqnarray}
\label{eq:disWmunu} W_{\mu \nu}^{V}(x_i, Q_i^2) = \left( -g_{\mu
\nu} + \frac{q_{i,\mu} q_{i,\nu}}{q_i^2} \right) F_{1}^{V}(x_i,
Q_i^2) + \frac{\hat{P}_{i,\mu} \hat{P}_{i,\nu}}{P_i \cdot q_i}
F_{2}^{V}(x_i, Q_i^2) + i \epsilon_{\mu \nu \alpha \beta}
\frac{P_i^\alpha q_i^\beta}{2 P_i \cdot q_i} F_{3}^{V}(x_i,
Q_i^2),~~ (V = Z, W^{\pm}),~
\end{eqnarray}
where $\epsilon_{\mu\nu\alpha\beta}$ is the completely antisymmetric
tensor and the momentum ${\hat P}_i$ is defined as
\begin{equation}
\label{eq:def-Phat} \hat{P}_{i,\mu} = P_{i,\mu} - \frac{P_i \cdot
q_i}{q_i^2} q_{i,\mu}.
\end{equation}
Due to the $CP$ conservation and the identity of the two final Higgs
bosons, the matrix element for the $W^-(-q_1) + W^+(-q_2)
\rightarrow h^0(k_1) + h^0(k_2)$ process is the same as that for the
$W^+(-q_1) + W^-(-q_2) \rightarrow h^0(k_1) + h^0(k_2)$ process,
i.e., ${\cal M}^{\mu\nu}_{W^-W^+} = {\cal M}^{\mu\nu}_{W^+W^-}$.
Here we depict the Feynman diagrams for the $ZZ \rightarrow h^0h^0$
and $W^+ W^- \rightarrow h^0h^0$ processes in Fig.\ref{fig:zzfusion}
and Fig.\ref{fig:wwfusion}, respectively, and the explicit
expressions for ${\cal M}^{\mu \nu}$ are presented in Appendix
\ref{sec:matrix-element}. Then the squared DIS hadronic tensor in
Eq.(\ref{eq:sfapproach}) can be written in the form as
\begin{eqnarray}
\label{eq:squaredht} W_{\mu \nu}^{V_1}(x_1, Q^2_1) {\cal M}^{\mu
\rho}_{V_1 V_2} {\cal M}^{\ast \nu \sigma}_{V_1 V_2} W_{\rho
\sigma}^{V_2}(x_2, Q^2_2) = \sum_{i,j=1}^3 C_{ij}^{V_1 V_2}
F_i^{V_1}(x_1, Q^2_1) F_j^{V_2}(x_2, Q^2_2),
\end{eqnarray}
where $C_{ij}^{V_1 V_2}$ can be automatically generated by using the
Mathematica packages FeynArts \cite{feynarts} and FeynCalc
\cite{feyncalc}.
\begin{figure}[ht!]
\centering
\includegraphics[scale=0.60]{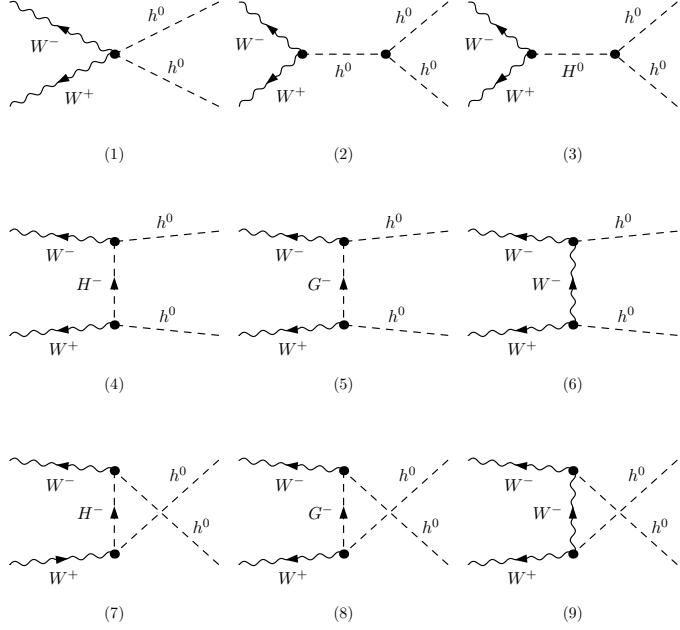}
\caption{Feynman diagrams for the $W^+ W^- \rightarrow h^0 h^0$ process.}
\label{fig:wwfusion}
\end{figure}

\par
Within the QCD factorization formalism, the structure functions can
be expressed as convolutions of the PDFs in proton with the short-distance
Wilson coefficient functions. We denote the gluon, quark and antiquark PDFs at the
factorization scale $\mu_f$ by $g(x, \mu_f)$, $q_i(x, \mu_f)$ and
$\bar{q}_i(x, \mu_f)$, respectively, where the subscript $i$
indicates the flavor of the (anti)quark. It is often convenient to
write the DIS structure functions in terms of the gluon, and the
following singlet and non-singlet quark distributions,
\begin{eqnarray}
&& ~q_{{\rm s}} = \sum_{i=1}^{n_f} \left( q_i + \bar{q}_i \right),
~~~~~~~~~~~~~~~~~~~~~~~~~~~~~~~~~~~~~~~~~~~~~~~  (\mbox{singlet}), \nonumber \\
&& q_{{\rm ns}}^{{\rm v}} = \sum_{i=1}^{n_f} \left( q_i - \bar{q}_i
\right),~~~ q^{\pm}_{{\rm ns}, ij} = \left( q_i \pm \bar{q}_i
\right) - \left( q_j \pm \bar{q}_j \right),~~~~~~
(\mbox{non-singlets}).~~~~~
\end{eqnarray}

\par
For the $Z$-exchange neutral current, the DIS structure functions
$F^Z_i~ (i = 1, 2, 3)$ can be written as follows \cite{big-vbf:prd}:
\begin{eqnarray}
\label{eq:DISF-Z}
F_{i}^{Z}(x, Q^2)
&=&
2 f_i(x)
\int_0^1 d y \int_0^1 d z \delta(x - y z) \sum_{j=1}^{n_f} \left( v_j^2 + a_j^2 \right) \nonumber \\
&&
\times
\Big[
q^{+}_{{\rm ns}, j}(y, \mu_f) C^{+}_{i, {\rm ns}}(z, Q, \mu_r, \mu_f)
+
q_{{\rm s}}(y, \mu_f) C_{i, {\rm q}}(z, Q, \mu_r, \mu_f)
+
g(y, \mu_f) C_{i, {\rm g}}(z, Q, \mu_r, \mu_f)
\Big], \nonumber \\
F_{3}^{Z}(x, Q^2)
&=&
2 f_3(x)
\int_0^1 d y \int_0^1 d z \delta(x - y z) \sum_{j=1}^{n_f} 2 v_j a_j \nonumber \\
&&
\times
\Big[
q^{-}_{{\rm ns}, j}(y, \mu_f) C^{-}_{3, {\rm ns}}(z, Q, \mu_r, \mu_f)
+
q_{{\rm ns}}^{{\rm v}}(y, \mu_f) C_{3, {\rm ns}}^{{\rm v}}(z, Q, \mu_r, \mu_f)
\Big],
\end{eqnarray}
where $i = 1, 2$, $f_1(x) = 1/2$, $f_2(x) = x$, $f_3(x) = 1$, and
the non-singlet quark densities $q^{\pm}_{{\rm ns}, i}$ are obtained
from $q^{\pm}_{{\rm ns}, ij}$ as
\begin{eqnarray}
q^{\pm}_{{\rm ns}, i} = \sum_{j = 1}^{n_f} q^{\pm}_{{\rm ns}, ij}~,~~~~~ (i = 1, ..., n_f).
\end{eqnarray}
The vector and axial-vector couplings of quark pair to $Z$ boson
used in Eqs.(\ref{eq:DISF-Z}) are given by
\begin{eqnarray}
v_i = I_i^3 - 2 Q_i \sin^2\theta_W,~~~~~~~~
a_i = I_i^3,
\end{eqnarray}
where $Q_i$ and $I_i^3$ are the electric charge and weak isospin of
the quark $q_i$, respectively.

\par
For the $W$-exchange charged current, the DIS structure functions
$F^{W^{\mp}}_i~ (i = 1, 2, 3)$ are expressed as follows:
\begin{eqnarray}
\label{eq:DISF-W}
F_{i}^{W^{\mp}}(x, Q^2)
&=&
f_i(x)
\int_0^1 d y \int_0^1 d z \delta(x - y z) \frac{1}{n_f} \sum_{j=1}^{n_f} \left( v_j^2 + a_j^2 \right) \nonumber \\
&&
\times
\Big[
\pm \delta q^{-}_{{\rm ns}}(y, \mu_f) C^{-}_{i, {\rm ns}}(z, Q, \mu_r, \mu_f)
+
q_{{\rm s}}(y, \mu_f) C_{i, {\rm q}}(z, Q, \mu_r, \mu_f)
+
g(y, \mu_f) C_{i, {\rm g}}(z, Q, \mu_r, \mu_f)
\Big], \nonumber \\
F_{3}^{W^{\mp}}(x, Q^2)
&=&
f_3(x)
\int_0^1 d y \int_0^1 d z \delta(x - y z) \frac{1}{n_f} \sum_{j=1}^{n_f} 2 v_j a_j \nonumber \\
&&
\times
\Big[
\pm \delta q^{+}_{{\rm ns}}(y, \mu_f) C^{+}_{3, {\rm ns}}(z, Q, \mu_r, \mu_f)
+
q_{{\rm ns}}^{{\rm v}}(y, \mu_f) C_{3, {\rm ns}}^{{\rm v}}(z, Q, \mu_r, \mu_f)
\Big],
\end{eqnarray}
where the non-singlet quark densities $\delta q^{\pm}_{{\rm ns}}$
are defined in terms of $q^{\pm}_{{\rm ns}, ij}$ as
\begin{eqnarray}
\delta q^{\pm}_{{\rm ns}} = \sum_{i \in {\rm up},~ j \in {\rm down}} q^{\pm}_{{\rm ns}, ij}~,
\end{eqnarray}
and the vector and axial-vector couplings for charged current are
given by
\begin{eqnarray}
v_i = a_i = \frac{1}{\sqrt{2}}.
\end{eqnarray}
We can see from Eqs.(\ref{eq:DISF-Z}) and Eqs.(\ref{eq:DISF-W}) that
the renormalization and factorization scales for quark densities in
each proton ($\mu_{1,r}$, $\mu_{2,r}$, $\mu_{1,f}$ and $\mu_{2,f}$)
enter in Eq.(\ref{eq:sfapproach}). The Wilson coefficient functions
in Eq.(\ref{eq:DISF-Z}) and Eq.(\ref{eq:DISF-W}) parameterize the
hard partonic scattering process and can be perturbatively expanded
in powers of $\alpha_s$. Up to the second order in $\alpha_s$,
$C_{3, {\rm ns}}^{{\rm v}} = C^{-}_{3, {\rm ns}}$, and the
perturbative expansion of these Wilson coefficient functions reads
\begin{eqnarray}
\label{eq:WF-NS-pm}
C^{\pm}_{i, {\rm ns}}
&=&
\delta(1-x)
+ a_s \biggl[ c_{i, {\rm ns}}^{(1), \pm} + L_M P_{{\rm ns}}^{(0), \pm} \biggr] \nonumber \\
&& +
\left.
a_s^2
\biggl[ c_{i, {\rm ns}}^{(2), \pm}
+ L_M
\biggl(
P_{{\rm ns}}^{(1), \pm} + c_{i, {\rm ns}}^{(1), \pm} \otimes (P_{{\rm ns}}^{(0), \pm} - \beta_0)
\biggr)
+ L_M^2
\biggl(
\frac{1}{2} P_{{\rm ns}}^{(0), \pm} \otimes (P_{{\rm ns}}^{(0), \pm} - \beta_0)
\biggr)
\right. \nonumber \\
&& +
\left.
\beta_0 L_R
\biggl( c_{i, {\rm ns}}^{(1), \pm} + L_M P_{{\rm ns}}^{(0), \pm} \biggr)
\biggr]
\right., ~~~~~~~~~~~~~~~~~~~~~~~~~~~~~~~~~~~~ (i = 1, 2, 3),
\end{eqnarray}
\begin{eqnarray}
\label{eq:WF-S-quark}
C_{i, {\rm q}}
&=&
\delta(1-x)
+ a_s \biggl[ c_{i, {\rm q}}^{(1)} + L_M P_{{\rm qq}}^{(0)} \biggr] \nonumber \\
&& +
\left.
a_s^2
\biggl[ c_{i, {\rm q}}^{(2)}
+ L_M
\biggl(
P_{{\rm qq}}^{(1)} + c_{i, {\rm q}}^{(1)} \otimes (P_{{\rm qq}}^{(0)} - \beta_0)
+ c_{i, {\rm g}}^{(1)} \otimes P_{{\rm gq}}^{(0)}
\biggr)
\right. \nonumber \\
&& +
\left.
L_M^2
\biggl(
\frac{1}{2} P_{{\rm qq}}^{(0)} \otimes (P_{{\rm qq}}^{(0)} - \beta_0)
+ \frac{1}{2} P_{{\rm qg}}^{(0)} \otimes P_{{\rm gq}}^{(0)}
\biggr)
\right. \nonumber \\
&& +
\left.
\beta_0 L_R
\biggl( c_{i, {\rm q}}^{(1)} + L_M P_{{\rm qq}}^{(0)} \biggr)
\biggr]
\right., ~~~~~~~~~~~~~~~~~~~~~~~~~ (i = 1, 2),
\end{eqnarray}
\begin{eqnarray}
\label{eq:WF-S-gluon}
C_{i, {\rm g}}
&=&
a_s \biggl[ c_{i, {\rm g}}^{(1)} + L_M P_{{\rm qg}}^{(0)} \biggr] \nonumber \\
&& +
\left.
a_s^2
\biggl[ c_{i, {\rm g}}^{(2)}
+ L_M
\biggl(
P_{{\rm qg}}^{(1)} + c_{i, {\rm q}}^{(1)} \otimes P_{{\rm qg}}^{(0)}
+ c_{i, {\rm g}}^{(1)} \otimes (P_{{\rm gg}}^{(0)} - \beta_0)
\biggr)
\right. \nonumber \\
&& +
\left.
L_M^2
\biggl(
\frac{1}{2} P_{{\rm qq}}^{(0)} \otimes P_{{\rm qg}}^{(0)}
+ \frac{1}{2} P_{{\rm qg}}^{(0)} \otimes (P_{{\rm gg}}^{(0)} - \beta_0)
\biggr)
\right. \nonumber \\
&& +
\left.
\beta_0 L_R
\biggl( c_{i, {\rm g}}^{(1)} + L_M P_{{\rm qg}}^{(0)} \biggr)
\biggr]
\right., ~~~~~~~~~~~~~~~~~~~~~~~~~ (i = 1, 2),
\end{eqnarray}
where $a_s = \alpha_s(\mu_r)/(4 \pi)$, $L_M = \ln(Q^2/\mu_f^2)$,
$L_R = \ln(\mu_r^2/\mu_f^2)$, $\beta_0 = (11 C_A - 2 n_f)/3$ is
referred to the one-loop beta-function coefficient, and $\otimes$
represents the standard Mellin convolution. It should be noted that
\footnote{For $i = 3$, the coefficient functions $c_{3, {\rm
ns}}^{(1), \pm}(x)$ satisfy $c_{3, {\rm ns}}^{(1), +}(x) = c_{3,
{\rm ns}}^{(1), -}(x)$ and are defined as $c_{3, {\rm q}}^{(1)}(x)$.
}
\begin{eqnarray}
P_{{\rm ns}}^{(0), \pm}
=
P_{{\rm qq}}^{(0)},~~~~~~~~~~~~~~~~~~~~~
c_{i, {\rm ns}}^{(1), \pm}
=
c_{i, {\rm q}}^{(1)},~~~~
(i = 1, 2, 3).
\end{eqnarray}
The two-loop order quark-quark splitting function $P_{{\rm
qq}}^{(1)}$ and the quark singlet DIS coefficient functions $c_{i,
{\rm q}}^{(2)}$ are usually expressed as
\begin{eqnarray}
P_{{\rm qq}}^{(1)} = P_{{\rm ns}}^{(1), +} + P_{{\rm ps}}^{(1)},~~~~~~~~~~~~~~~~~~~~~
c_{i, {\rm q}}^{(2)} = c_{i, {\rm ns}}^{(2), +} + c_{i, {\rm ps}}^{(2)},~~~~(i = 1, 2),~~
\end{eqnarray}
where $P_{{\rm ps}}^{(1)}$ and $c_{i, {\rm ps}}^{(2)}$ are the
pure-singlet contributions at the second order of $\alpha_s$. All
the DIS coefficient functions $c_{i, {\rm q}}^{(1)}$, $c_{3, {\rm
q}}^{(1)}$, $c_{i, {\rm g}}^{(1)}$, $c_{i, {\rm ns}}^{(2), \pm}$,
$c_{i, {\rm ps}}^{(2)}$, $c_{i, {\rm g}}^{(2)}$ $(i = 1, 2)$ and the
splitting functions $P_{{\rm qq}}^{(0)}$, $P_{{\rm qg}}^{(0)}$,
$P_{{\rm gq}}^{(0)}$, $P_{{\rm gg}}^{(0)}$, $P_{{\rm ns}}^{(1),
\pm}$, $P_{{\rm ps}}^{(1)}$, $P_{{\rm qg}}^{(1)}$ used in
Eqs.(\ref{eq:WF-NS-pm})-(\ref{eq:WF-S-gluon}) are given in
Refs.\cite{vanNeerven:1999ns, vanNeerven:1999distl,
vanNeerven:2000uj, Vermaseren:2005qc,Moch:2008fj}. They can be
easily evaluated in terms of harmonic polylogarithms
$H_{\vec{m}}(x)/(1 \pm x)$ \cite{Remiddi:1999ew}. In this paper we
adopt the {\sc Fortran} program Hplog \cite{Gehrmann} to implement
numerical calculation of harmonic polylogarithms.

\vskip 5mm
\section{Numerical results and discussion}
\label{sec:NumericalResults}
\par
In this section we present the integrated cross sections and some
kinematic distributions for the light $CP$-even Higgs pair
production via VBF at $\sqrt{S} = 14$, $33$ and $100~ {\rm TeV}$
proton-proton colliders up to the QCD NNLO by employing the SF
approach. In our numerical calculations we use the following values
for the electroweak parameters:
\begin{eqnarray}
M_W = 80.385~ {\rm GeV},~~~~~
M_Z = 91.1876~ {\rm GeV},~~~~~
G_F = 1.1663787 \times 10^{-5}~ {\rm GeV}^{-2}.
\end{eqnarray}
The Weinberg angle is fixed in the on-shell scheme as
$\sin^2\theta_W = 1 - M_W^2/M_Z^2$. We choose the 2HDM(II) input
parameters at two benchmark points, B1 and B2, for demonstration and
comparison, whose related parameters are listed in Table
\ref{tab:benchmark}. The parameters at both the B1 and B2 points
survive in the present theoretical and experimental constraints
\cite{sushufang:2hdm}. At the benchmark point B1 we have $m_{H^0}>2
m_{h^0}$ and there exists $H^0$ resonance effect in the VBF
$h^0$-pair production process. While at the benchmark point B2 there
does not exist $H^0$ resonance effect, and the corresponding results
should be the same with those in the SM case for the VBF $h^0 h^0 +
2~ jets$ production process. The width of $H^0$ can be calculated by
using 2HDMC program \cite{cpc:2hdmc}, and at the benchmark point B1
we get the total decay width of $H^0$ boson being
$\Gamma_{H^0}=5.484~{\rm GeV}$.
\begin{table}
  \centering
  \begin{tabular}{c|c|c|c|c|c|c}
    \hline\hline
     ~~~~~ & ~$\sin(\beta-\alpha)$ & ~~~~$\tan\beta$~~~~ & $m_{h^0}$~(GeV) & $m_{H^0}$~(GeV)
     & $m_{A^0}$~(GeV) & $m_{H^{\pm}}$~(GeV) \\
    \hline
    B1 & 0.6 & 2 & 126 & 275 & 600 & 600 \\
    B2 & 1 & 1.5 & 126 & 160 & 380 & 420 \\
    \hline\hline
  \end{tabular}
\caption{The 2HDM benchmark points.} \label{tab:benchmark}
\end{table}

\par
We adopt the MSTW2008 PDFs \cite{mstw2008} in the convolutions of
parton densities with Wilson coefficient functions. In the
calculations of the $ZZ$-fusion contributions (see
Fig.\ref{fig:zzfusion}), we take the $b$-quark as a massless parton
and the number of massless flavors $n_f=5$ in Eqs.(\ref{eq:DISF-Z}).
While in the evaluations of the $WW$-fusion process (see
Fig.\ref{fig:wwfusion}), the initial $b$-quark is not included since
it would produce a top-quark in the final state. In the following
analysis we take $\mu = \mu_f = \mu_r$ for simplicity and the
typical central value of the renormalization/factorization scale is
fixed by the corresponding vector-boson momentum transfer
$\mu^2=-q_i^2=Q^2$ for $i=1,2$ \footnote{Here the scale $\mu^2=Q^2$
means $\mu_1^2=\mu_{1,f}^2=\mu_{1,r}^2=Q_1^2$ and
$\mu_2^2=\mu_{2,f}^2=\mu_{2,r}^2=Q_2^2$. }, if there is no other
statement. Furthermore, we put a lower bound of $Q^2 > 4~{\rm
GeV}^2$ in order to keep in the perturbative regime, and the
independence of the integrated cross section on this technical $Q$
cut has been checked numerically.

\vskip 5mm
\subsection{Dependence on 2HDM(II) parameters }
\label{sec:TheoUncertainty}
\par
The integrated cross section for the VBF light, neutral $CP$-even
Higgs boson pair production is related to the 2HDM(II) parameters,
such as the two $CP$-even Higgs boson masses, ratio of the VEVs and
the mixing angle between the two $CP$-even Higgs bosons. In this
subsection we study the dependence of integrated cross section for
the VBF $h^0h^0+2~jets$ production on the related model parameters
at the $\sqrt{S}=14~{\rm TeV}$ LHC by adopting above event selection
scheme.

\par
\begin{figure}[ht!]
\centering
\includegraphics[scale=0.50]{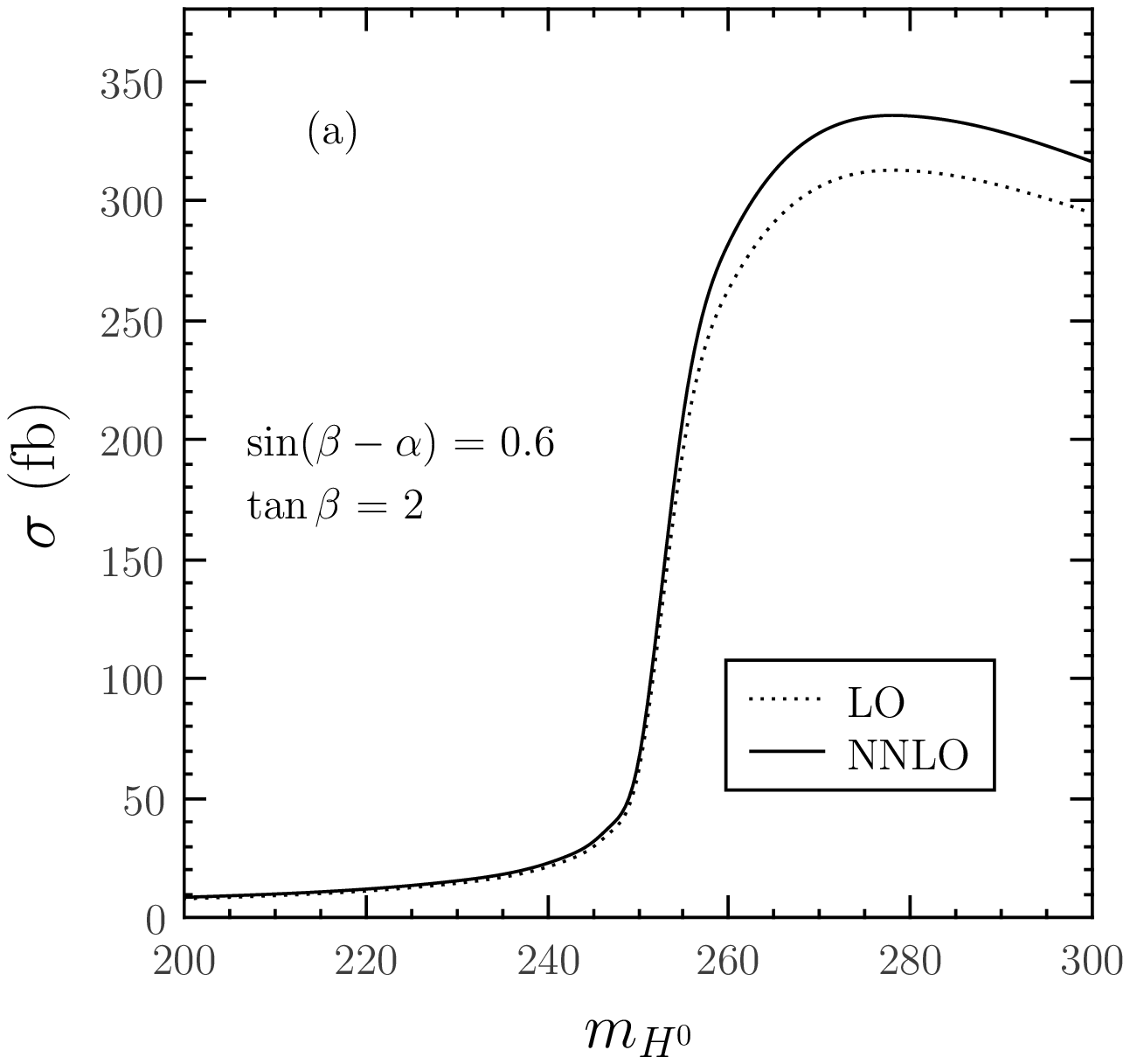}
\includegraphics[scale=0.50]{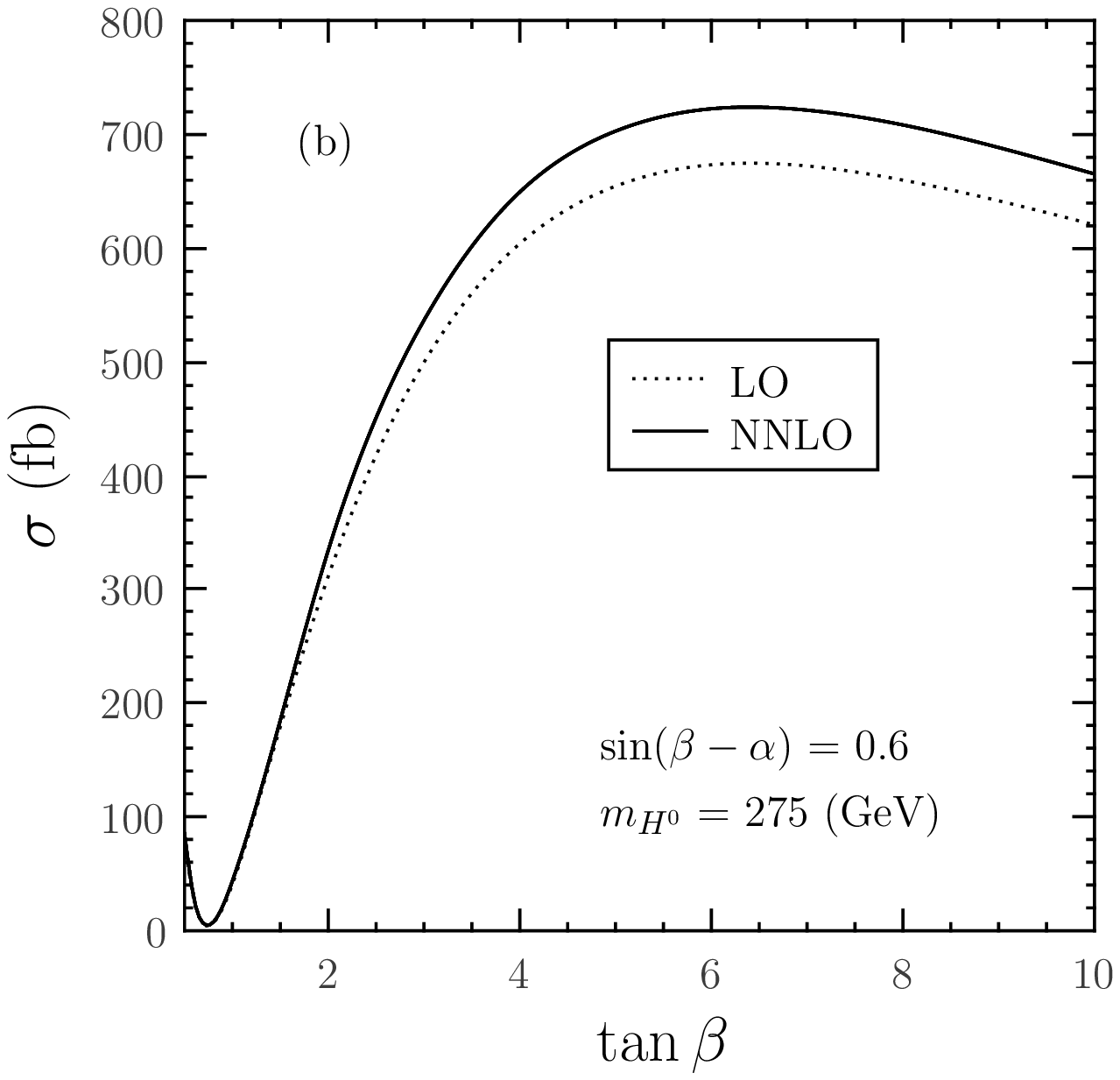}
\includegraphics[scale=0.50]{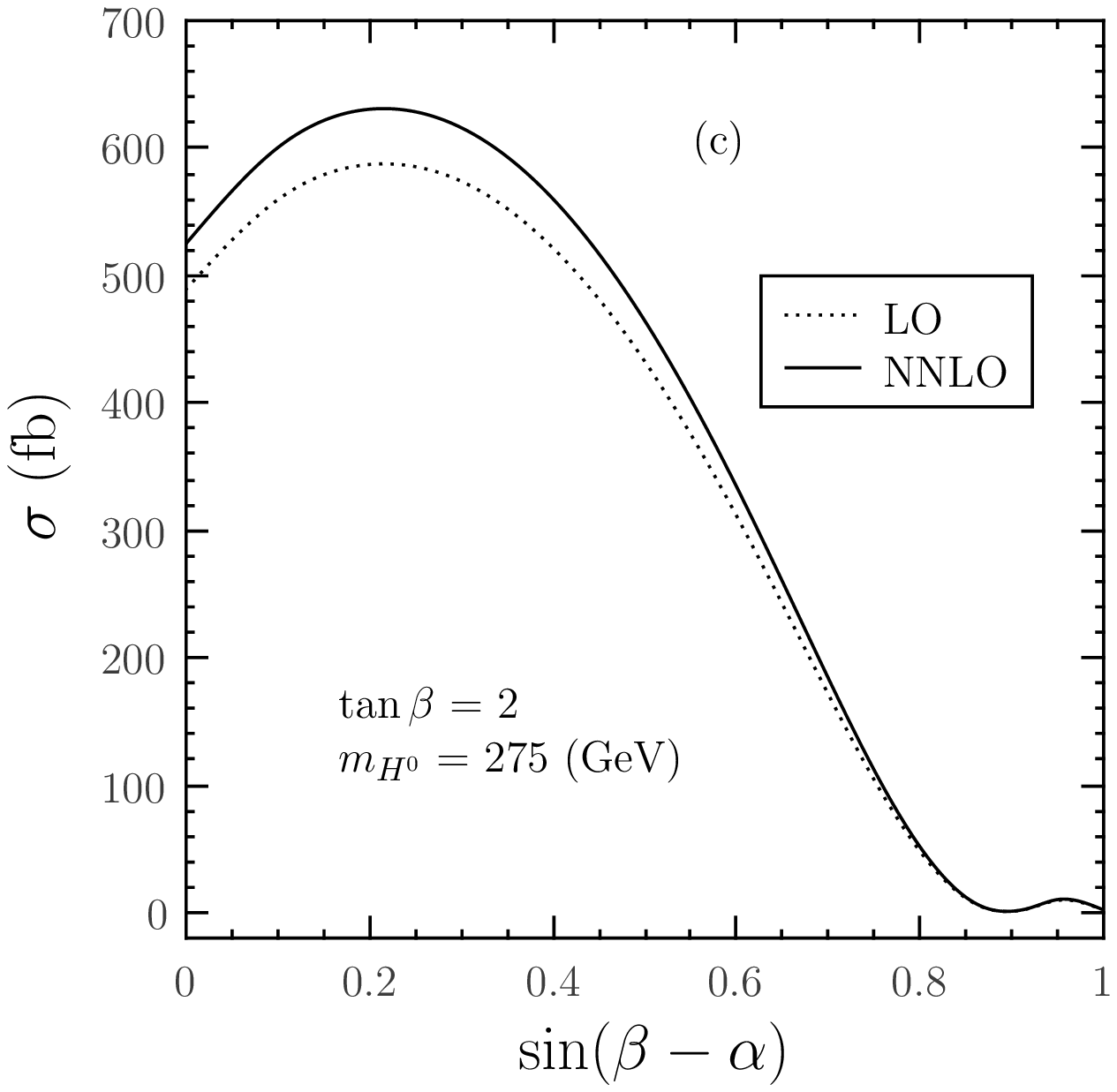}
\caption{The dependence of the LO and NNLO QCD corrected integrated
cross sections for the VBF $h^0 h^0 + 2~ jets$ production on the
2HDM(II) parameters at the $\sqrt{S} = 14~ {\rm TeV}$ LHC. (a) as
the function of the $H^0$ boson mass, (b) as the function of the
$\tan\beta$, (c) as the function of the parameter
$\sin(\beta-\alpha)$. } \label{fig:2HDM-parameter}
\end{figure}

In Fig.\ref{fig:2HDM-parameter}(a) we depict the LO and NNLO QCD
corrected integrated cross sections as functions of $m_{H^0}$ with
the other related model parameters being the values at the benchmark
point B1. We see from the figure that there is a steep increment at
the position of $m_{H^0} \sim 2 m_{h^0} =252~{\rm GeV}$ due to the
on-shell $H^0$ decay of $H^0 \to h^0h^0$, and $H^0$ resonance effect
enhances the production rate obviously in the region of $m_{H^0} >
260~{\rm GeV}$. It shows also that the QCD corrections up to NNLO
always increase the LO cross section particularly for the large
$H^0$ mass.

\par
Fig.\ref{fig:2HDM-parameter}(b) shows the dependence of the LO and
QCD NNLO corrected integrated cross sections on the ratio of the
VEVs $\tan\beta$. There we fix all the 2HDM(II) parameters are the
values of the benchmark point B1 except $\tan\beta$, which varies
from $0.5$ to $10$. The figure demonstrates that both the LO and
NNLO corrected total cross sections reach their minimal and maximal
values at the positions about $\tan\beta \sim 0.75$ and $6.0$,
respectively. In the region of $\tan\beta > 4.0$, both the LO and
the NNLO QCD corrected cross sections exceed $600~fb$.

\par
We plot Fig.\ref{fig:2HDM-parameter}(c) to show the dependence of
the LO and NNLO QCD corrected integrated cross sections on the
parameter $\sin(\beta-\alpha)$ with the other related 2HDM(II)
parameters being fixed at the benchmark point B1, i.e.,
$\tan\beta=2$ and $m_{H^0} = 275~ {\rm GeV}$. It shows obviously
that both the LO and the NNLO QCD corrected total cross sections
reach their maxima at the position of $\sin(\beta-\alpha)=0.2$, and
then decrease with the increment of $\sin(\beta-\alpha)$ from $0.2$
to $0.9$.

\vskip 5mm
\subsection{Theoretical uncertainties of integrated cross section }
\label{sec:TheoUncertainty}
\par
In order to make a precision comparison between the theoretical
predictions and experimental measurements, we should assess
thoroughly the theoretical uncertainties affecting the central
predictions of the total cross sections. For some production
processes at hadron colliders, such as $pp \to V^*V^* \to h^0 h^0 +
2~ jets$ process, the theoretical uncertainty mainly comes from the
missing higher order corrections, PDFs and $\alpha_s$.

\vskip 5mm
\subsubsection{Scale uncertainty}
\par
\begin{figure}[ht!]
\centering
\includegraphics[scale=0.50]{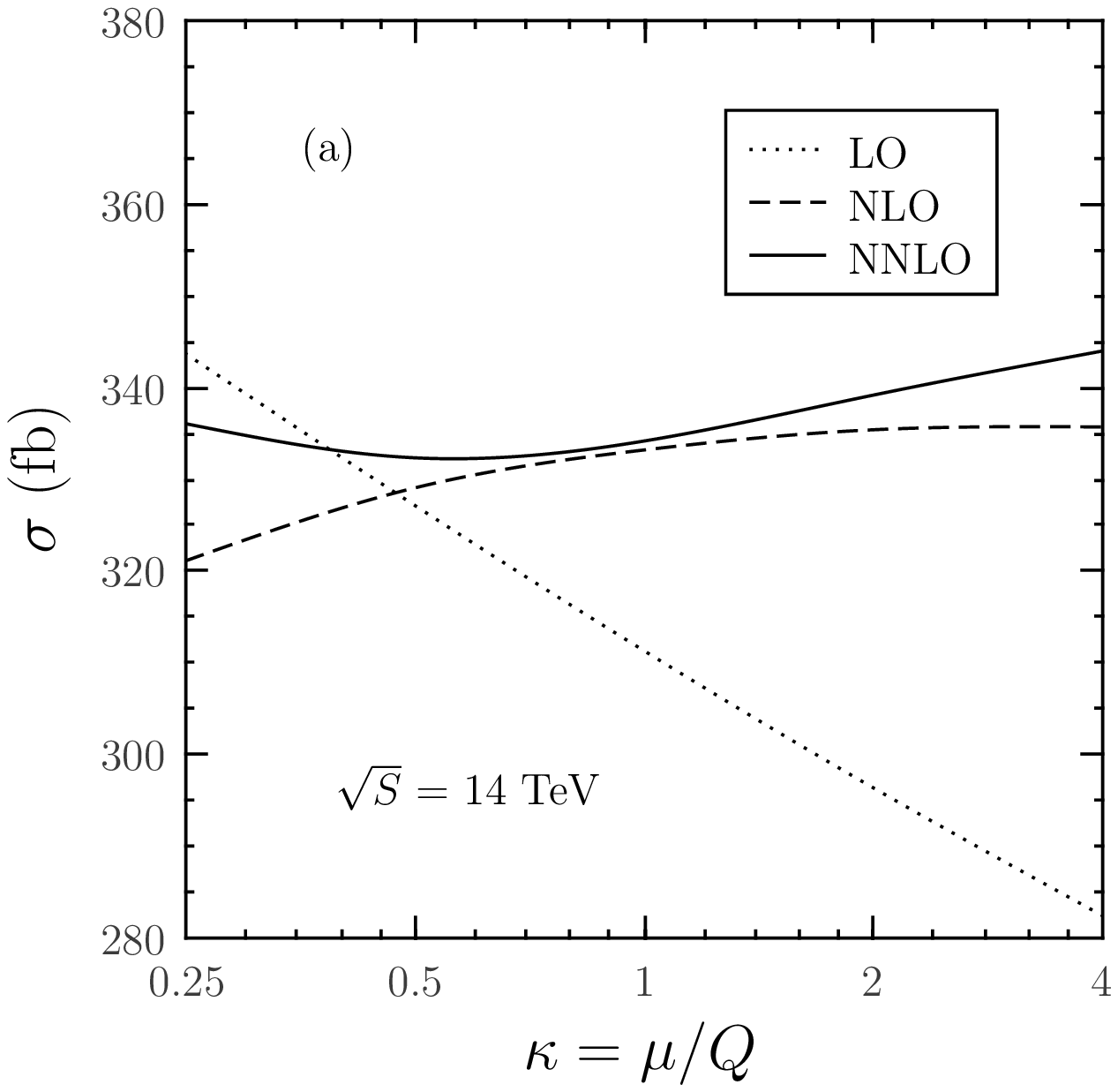}
\includegraphics[scale=0.50]{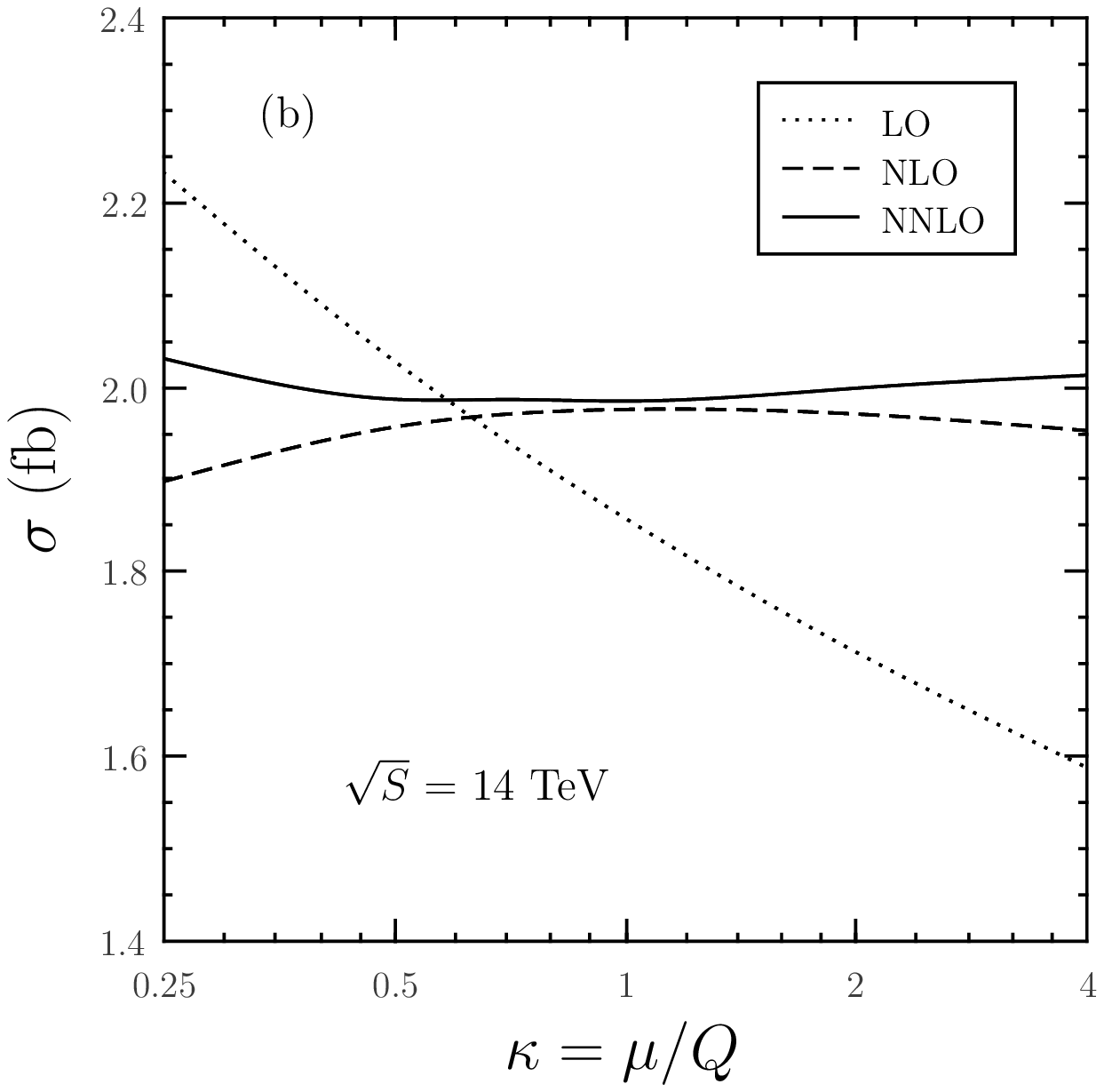}
\caption{The scale dependence of the LO, QCD NLO and NNLO corrected
integrated cross sections for the VBF $h^0 h^0 + 2~ jets$ production
at the $\sqrt{S} = 14~ {\rm TeV}$ LHC. (a) at the benchmark point
B1. (b) at the benchmark point B2. } \label{fig:x-scale}
\end{figure}

The uncertainty due to missing higher order radiative corrections
can be estimated by varying the factorization/renormalization scale
$\mu$ around a central value that is taken close to the physical
scale of the process. A conventional range of variation for the VBF
process is
\begin{eqnarray}
\frac{1}{4} Q \leq \mu \leq 4 Q,
\end{eqnarray}
where the central value $Q$ of $\mu_r$ and $\mu_f$ is the virtuality
of the vector bosons which fuse into the Higgs boson pair. In
Figs.\ref{fig:x-scale}(a) and (b) we present the scale dependence of
the LO, QCD NLO and NNLO corrected integrated cross sections for the
VBF $h^0 h^0 + 2~ jets$ production at the $\sqrt{S} = 14~ {\rm TeV}$
LHC at the benchmark points B1 and B2, respectively. The central
values of the integrated cross sections and the corresponding errors
due to missing higher order radiative corrections are listed in
Table \ref{tab:x-error}. From Figs.\ref{fig:x-scale}(a, b) and Table
\ref{tab:x-error} we find that the scale uncertainties of integrated
cross sections can be significantly reduced by including higher
order radiative corrections. For the benchmark B1 (B2), the
corresponding relative upper and lower scale relative uncertainties,
defined as: ${\rm the~upper~limit~ of~scale~uncertainty}\equiv
\frac{max \left[ \sigma(\mu) - \sigma(\mu=Q)
\right]}{\sigma(\mu=Q)}$, and ${\rm
the~lower~limit~of~scale~uncertainty}\equiv \frac{min \left[
\sigma(\mu) - \sigma(\mu=Q) \right]}{\sigma(\mu=Q)}$ with $\mu \in
[Q/4,~ 4 Q]$, are about $^{(+10\%)}_{(-9\%)}$
$\left(^{(+20\%)}_{(-15\%)}\right)$ at the LO, and are reduced to
$^{(+0.8\%)}_{(-3.6\%)}$ $\left(^{(+0.0\%)}_{(-3.9\%)}\right)$ and
$^{(+2.9\%)}_{(-0.5\%)}$ $\left(^{(+2.3\%)}_{(-0.0\%)}\right)$ at the
QCD NLO and NNLO, respectively. We see that the variation of scale
uncertainty of $\sigma_{NNLO}$ is smaller than the corresponding
ones of $\sigma_{LO}$ and $\sigma_{NLO}$. Therefore, from the point
of view of improving the scale uncertainty, the NNLO QCD corrections
should be taken into account for the precision measurement of the
VBF Higgs pair production process.
\begin{table}
  \centering
  \begin{tabular}{c|c|c|c}
   \hline\hline
     ~~~~~ & $\sigma_{LO}$ (fb) & $\sigma_{NLO}$ (fb) & $\sigma_{NNLO}$ (fb) \\
    \hline
    B1 &   ~$311.30^{+32.38~(+10\%)}_{-28.88~(-9\%)}$~     &  ~$333.20^{+2.51~(+0.8\%)}_{-12.08~(-3.6\%)}$~   &   ~$334.18^{+9.82~(+2.9\%)}_{-1.83~(-0.5\%)}$~ \\
    B2 &   ~$1.858^{+0.374~(+20\%)}_{-0.270~(-15\%)}$~     &  ~$1.976^{+0.00~~(+0.0\%)}_{-0.078~(-3.9\%)}$~   &   ~$1.986^{+0.045~(+2.3\%)}_{-0.00~(-0.0\%)}$ ~ \\
    \hline\hline
  \end{tabular}

\caption{The LO, QCD NLO and NNLO corrected integrated cross
sections for the VBF $h^0 h^0 + 2~ jets$ production at $\sqrt{S} =
14~ {\rm TeV}$ LHC at the benchmark points B1 and B2. The scale
uncertainties are obtained from the variation in the range of $\mu
\in [Q/4,~ 4 Q]$. The data in brackets are the relative
uncertainties. } \label{tab:x-error}
\end{table}

\par
Figs.\ref{fig:x-scale}(a) and (b) also demonstrate that the impact
of the NNLO QCD corrections at the central scale $Q$ is smaller than
$1\%$. Compared with other values in the range $[Q/4, 4Q]$, the
scale choice $\mu = Q$ is more natural because it exhibits a better
convergence of the perturbative expansion. Therefore, we set the
factorization/renormalization scale $\mu$ as its central value $Q$
in the following numerical calculations.

\vskip 5mm
\subsubsection{PDF+$\alpha_s$ uncertainty}
\par
For a given parametrization of the PDFs such as MSTW2008, the PDF
uncertainty comes from the experimental uncertainties on the fitted
data. For a fixed value of $\alpha_s$, MSTW2008 provides a central
PDF set $S_0$ and $2 n$ eigenvector PDF sets $S^{\pm}_i~ (i = 1,
..., n,~ n = 20)$. The PDF uncertainties on the hadronic cross
section are given by \cite{mstw2008:method}
\begin{eqnarray}
\label{eq:pdferror-mstw}
\left( \Delta \sigma^{\alpha_s}_{PDF} \right)_{+}
&=&
\sqrt{
\sum_{i=1}^{n}
\biggl\{
\max
\Big[
\sigma^{\alpha_s}(S^+_i) - \sigma^{\alpha_s}(S_0),~
\sigma^{\alpha_s}(S^-_i) - \sigma^{\alpha_s}(S_0),~
0
\Big]
\biggr\}^2
}, \nonumber \\
\left( \Delta \sigma^{\alpha_s}_{PDF} \right)_{-}
&=&
\sqrt{
\sum_{i=1}^{n}
\biggl\{
\max
\Big[
\sigma^{\alpha_s}(S_0) - \sigma^{\alpha_s}(S^+_i),~
\sigma^{\alpha_s}(S_0) - \sigma^{\alpha_s}(S^-_i),~
0
\Big]
\biggr\}^2
},
\end{eqnarray}
where $\sigma^{\alpha_s}(S_0)$, $\sigma^{\alpha_s}(S_i^{+})$ and
$\sigma^{\alpha_s}(S_i^{-})$ represent the cross sections obtained
by using the PDF sets $S_0$, $S_i^{+}$ and $S_i^{-}$, respectively.

\par
In additional to the PDF uncertainty, there is also an uncertainty
due to the errors on the value of the strong coupling constant
$\alpha_s$. Beside the best-fit sets of PDFs which correspond to
$\alpha_s^0$, four more PDF sets corresponding to $\alpha_s =
\alpha_s^0 \pm 0.5 \sigma$ and $\alpha_s = \alpha_s^0 \pm 1 \sigma$
are provided by the MSTW collaboration, where $\alpha_s^0$ and
$\sigma$ are the central value and the standard deviation of
$\alpha_s$, respectively. Comparing the results obtained from the
five sets, the pure $\alpha_s$ uncertainties are defined as
\begin{eqnarray}
\label{eq:aserror-mstw}
\left( \Delta \sigma_{\alpha_s} \right)_{+}
&=&
\max_{\alpha_s}
\Big[
\sigma^{\alpha_s}(S_0)
\Big]
- \sigma^{\alpha_s^0}(S_0), \nonumber \\
\left( \Delta \sigma_{\alpha_s} \right)_{-}
&=&
\sigma^{\alpha_s^0}(S_0) -
\min_{\alpha_s}
\Big[
\sigma^{\alpha_s}(S_0)
\Big],
\end{eqnarray}
where max and min run over the five values of $\alpha_s$.

\par
For the MSTW2008 PDFs, the combined PDF+$\alpha_s$ uncertainties are
given by \cite{mstw2008:method}
\begin{eqnarray}
\label{eq:pdfaserror-mstw}
\left( \Delta \sigma_{PDF+\alpha_s} \right)_{+}
&=&
\max_{\alpha_s}
\Big[
\sigma^{\alpha_s}(S_0) + \left( \Delta \sigma^{\alpha_s}_{PDF} \right)_{+}
\Big]
- \sigma^{\alpha_s^0}(S_0), \nonumber \\
\left( \Delta \sigma_{PDF+\alpha_s} \right)_{-}
&=&
\sigma^{\alpha_s^0}(S_0) -
\min_{\alpha_s}
\Big[
\sigma^{\alpha_s}(S_0) - \left( \Delta \sigma^{\alpha_s}_{PDF} \right)_{-}
\Big].
\end{eqnarray}
If the dependence of $\left( \Delta \sigma^{\alpha_s}_{PDF}
\right)_{\pm}$ on $\alpha_s$ is negligible, the overall
PDF+$\alpha_s$ uncertainties can be approximately expressed as
\begin{eqnarray}
\label{eq:pdfaserror-appro}
\left( \Delta \sigma_{PDF+\alpha_s} \right)_{\pm}
&\simeq&
\left( \Delta \sigma_{PDF} \right)_{\pm} +
\left( \Delta \sigma_{\alpha_s} \right)_{\pm},
\end{eqnarray}
where $\left( \Delta \sigma_{PDF} \right)_{\pm} = (\Delta
\sigma^{\alpha_s^0}_{PDF})_{\pm}$ are pure PDF uncertainties. In the
following calculations, we adopt Eq.(\ref{eq:pdfaserror-appro}) to
evaluate the combined PDF+$\alpha_s$ uncertainty.

\vskip 5mm
\subsubsection{Integrated cross sections}
\par
In Table \ref{tab:x-ppcolliders} we present the LO, QCD NLO and NNLO
corrected integrated cross sections for the VBF $h^0 h^0 + 2~ jets$
production at $\sqrt{S} = 14$, $33$ and $100~ {\rm TeV}$ $pp$ hadron
colliders at the benchmark points B1 and B2. The scale and combined
PDF+$\alpha_s$ uncertainties are also provided to estimate the
precisions of these perturbative predictions. From this table we can
see that the factorization/renormalization scale and the combined
PDF+$\alpha_s$ uncertainties are generally comparable, and both of
them are reduced by NLO, NNLO QCD corrections. For both benchmarks
of B1 and B2, the theoretical upper and lower deviations of the NNLO
prediction at the $14~{\rm TeV}$ LHC, which are obtained by adding
linearly the scale and $PDF+\alpha_s$ uncertainties, are always
bellow $5.1\%$. As the increment of $pp$ colliding energy $\sqrt{S}$
from $14~ {\rm TeV}$ to $100~ {\rm TeV}$, the NNLO QCD corrections
increase the integrated cross sections and the combined
uncertainties for the VBF Higgs pair production at the benchmarks B1
and B2 separately.
\begin{table}
  \centering \small
  \begin{tabular}{c|c|c|c|c}
  \hline\hline
  \multicolumn{2}{c|}{$\sqrt{S}$ (TeV)} & $\sigma_{LO}$ (fb) & $\sigma_{NLO}$ (fb) & $\sigma_{NNLO}$ (fb) \\
  \hline
  \multirow{3}{*}{14}  & B1 &
   $311.30^{+32.38 +4.06}_{-28.88-4.04} \left( ^{+ 10.4 \% + 1.3 \%}_{-9.3 \% - 1.3 \%} \right)$ &
   $333.20^{+2.51 +8.46}_{-12.08-6.57} \left( ^{+ 0.8 \% + 2.5 \%}_{-3.6 \% - 2.0 \%} \right)$ &
   $334.18^{+9.82 +7.36}_{-1.83-5.87} \left( ^{+ 2.9 \% + 2.2 \%}_{-0.5 \% - 1.8 \%} \right)$ \\
   & B2 &
   $1.858^{+0.374 +0.028}_{-0.270-0.026} \left( ^{+ 20.1 \% + 1.5 \%}_{-14.5 \% - 1.4 \%} \right)$ &
   $1.976^{+0 +0.052}_{-0.078-0.039} \left( ^{+ 0 \% + 2.6 \%}_{-3.9 \% - 2.0 \%} \right)$ &
   $1.986^{+0.045 +0.048}_{-0-0.035} \left( ^{+ 2.3 \% + 2.4 \%}_{-0.0 \% - 1.8 \%} \right)$ \\
   \hline
   \multirow{3}{*}{33}  & B1 &
   $1404^{+0 +15}_{-30-16} \left( ^{+ 0 \% + 1.1 \%}_{-2.1 \% - 1.1 \%} \right)$ &
   $1500^{+54 +35}_{-74-32} \left( ^{+ 3.6 \% + 2.3 \%}_{-4.9 \% - 2.1 \%} \right)$ &
   $1503^{+73 +32}_{-17-28} \left( ^{+ 4.9 \% + 2.1 \%}_{-1.1 \% - 1.9 \%} \right)$ \\
   & B2 &
   $11.234^{+0.878 +0.129}_{-0.830-0.149} \left( ^{+ 7.8 \% + 1.1 \%}_{-7.4 \% - 1.3 \%} \right)$ &
   $12.002^{+0.190 +0.297}_{-0.562-0.225} \left( ^{+ 1.6 \% + 2.5 \%}_{-4.7 \% - 1.9 \%} \right)$ &
   $12.041^{+0.359 +0.258}_{-0.060-0.209} \left( ^{+ 3.0 \% + 2.1 \%}_{-0.5 \% - 1.7 \%} \right)$ \\
   \hline
   \multirow{3}{*}{100}  & B1 &
   $7271^{+770 +73}_{-1130-81} \left( ^{+ 10.6 \% + 1.0 \%}_{-15.5 \% - 1.1 \%} \right)$ &
   $7554^{+535 +188}_{-580-119} \left( ^{+ 7.1 \% + 2.5 \%}_{-7.7 \% - 1.6 \%} \right)$ &
   $7578^{+553 +150}_{-134-170} \left( ^{+ 7.3 \% + 2.0 \%}_{-1.8 \% - 2.2 \%} \right)$ \\
   & B2 &
   $75.36^{+4.91 +2.07}_{-6.34-1.07} \left( ^{+ 6.5 \% + 2.7 \%}_{-8.4 \% - 1.4 \%} \right)$ &
   $79.82^{+3.92 +2.99}_{-5.26-1.95} \left( ^{+ 4.9 \% + 3.7 \%}_{-6.6 \% - 2.4 \%} \right)$ &
   $80.05^{+3.92 +1.58}_{-0.80-1.48} \left( ^{+ 4.9 \% + 2.0 \%}_{-1.0 \% - 1.8 \%} \right)$ \\
   \hline \hline
  \end{tabular}
\caption{The LO, QCD NLO and NNLO corrected integrated cross
sections for VBF $h^0 h^0 + 2~ jets$ production at $\sqrt{S} = 14$,
33 and $100~ {\rm TeV}$ $pp$ colliders at the benchmark points B1
and B2 together with scale uncertainties (the first ones) and
combined $68\%$ CL PDF+$\alpha_s$ uncertainties (the second ones).
The data in brackets are the relative uncertainties. }
\label{tab:x-ppcolliders}
\end{table}

\vskip 5mm
\subsection{Kinematic distributions}
\label{sec:kinematicalDis}
\par
Analogous to the VBF $h^0 + 2~ jets$ production, the signal of VBF
$h^0 h^0 + 2~ jets$ production involves two energetic forward and
backward jets in association with two centrally produced Higgs
bosons. This character plays an important role in discriminating the
VBF signal from the heavy QCD background. Since a precision study of
the kinematic distributions of the final Higgs bosons for the signal
process is helpful in theoretical and experimental analyses, we
provide the NNLO QCD corrected transverse momentum $p_T$, rapidity
$y$ and invariant mass $M$ distributions of the final Higgs bosons
for the VBF $h^0 h^0 + 2~ jets$ production at $pp$ colliders. In
order to assess the impact of the NNLO QCD corrections, we introduce
the differential NNLO QCD $K$-factor, which is defined as
\begin{eqnarray}
K(x) = \frac{d \sigma_{NNLO}}{dx}\Big/\frac{d \sigma_{LO}}{dx},
\end{eqnarray}
where $x$ stands for a kinematic variable.

\par
\begin{figure}[ht!]
\centering
\includegraphics[scale=0.50]{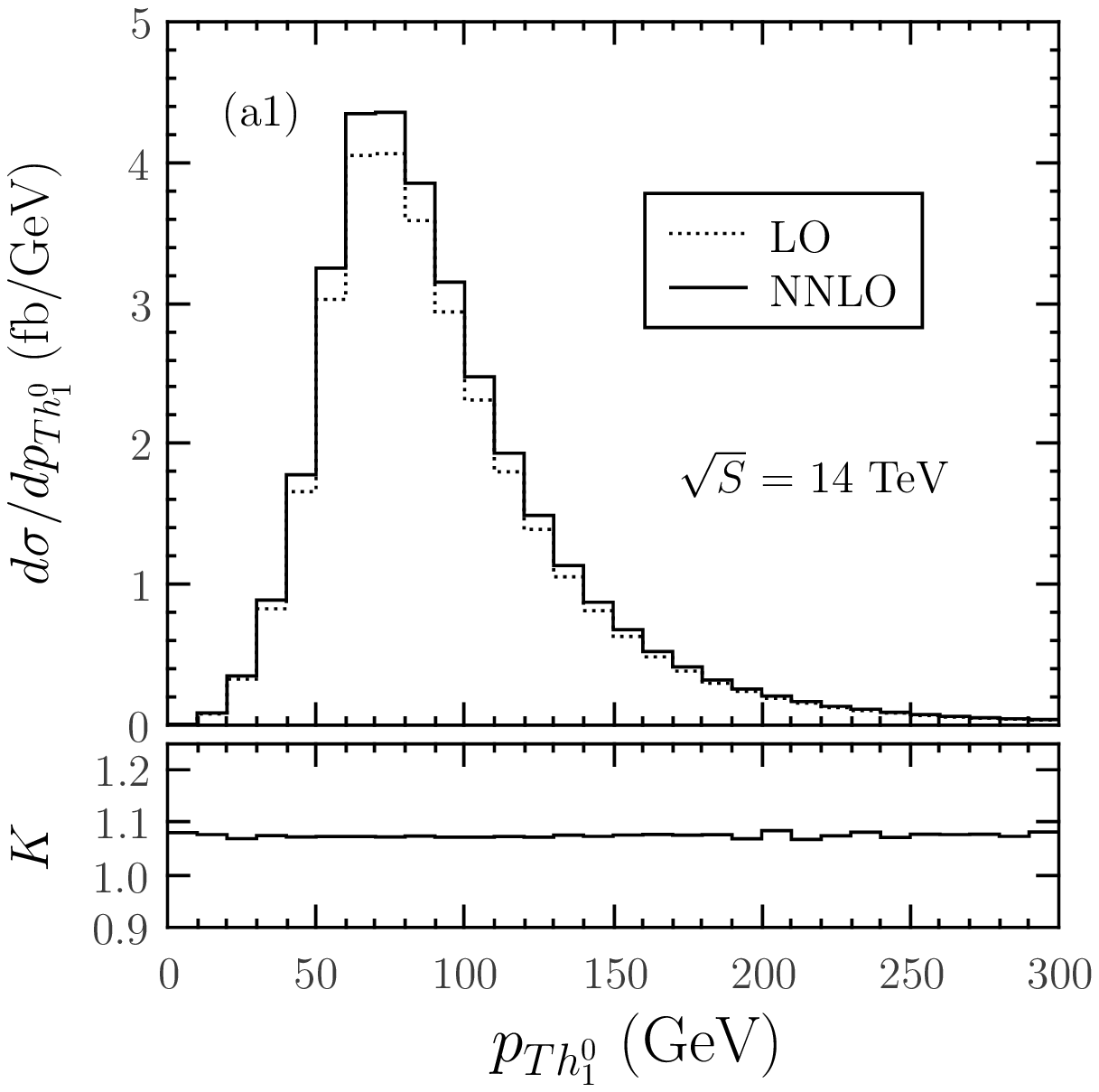}
\includegraphics[scale=0.50]{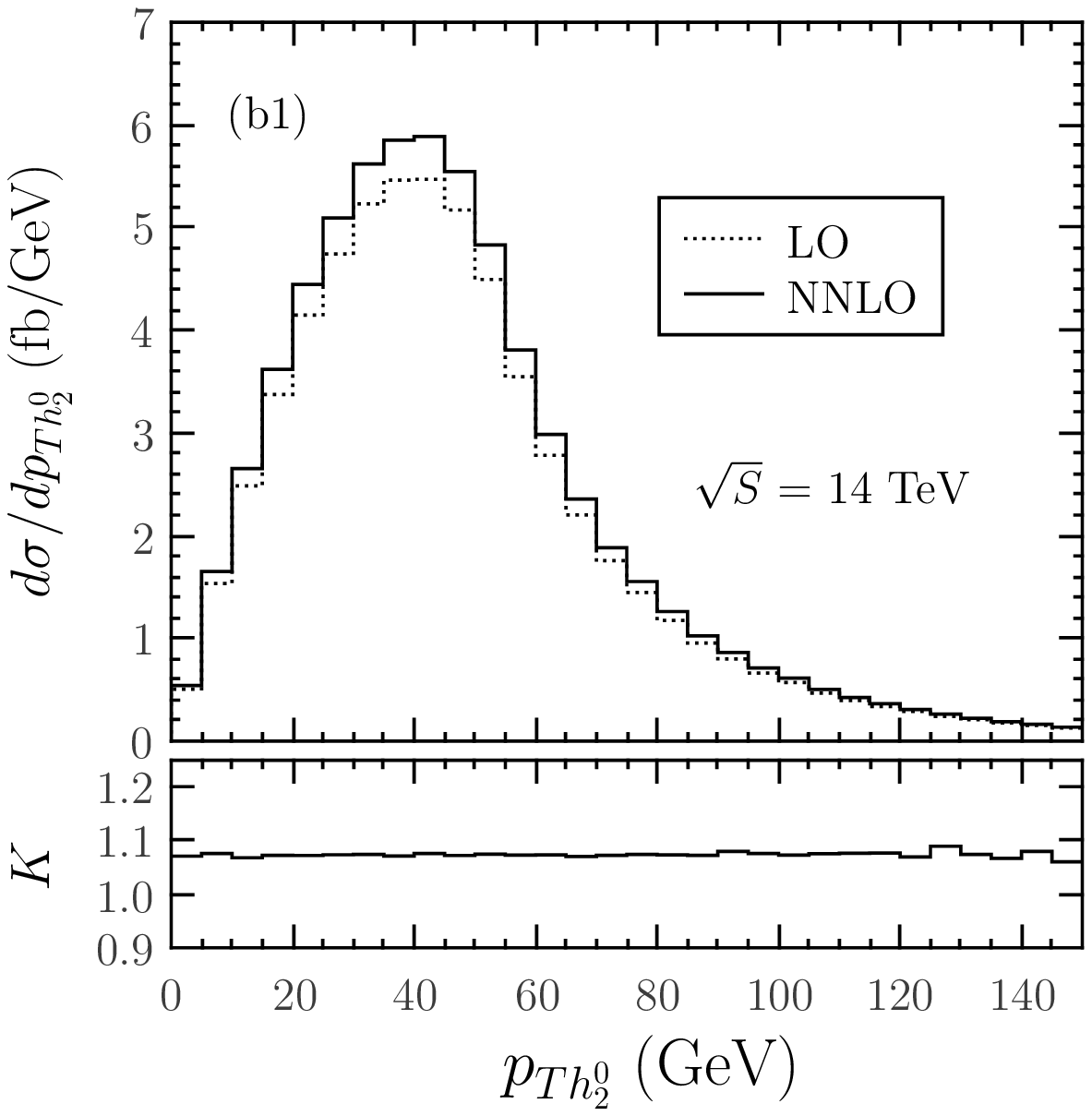}
\includegraphics[scale=0.50]{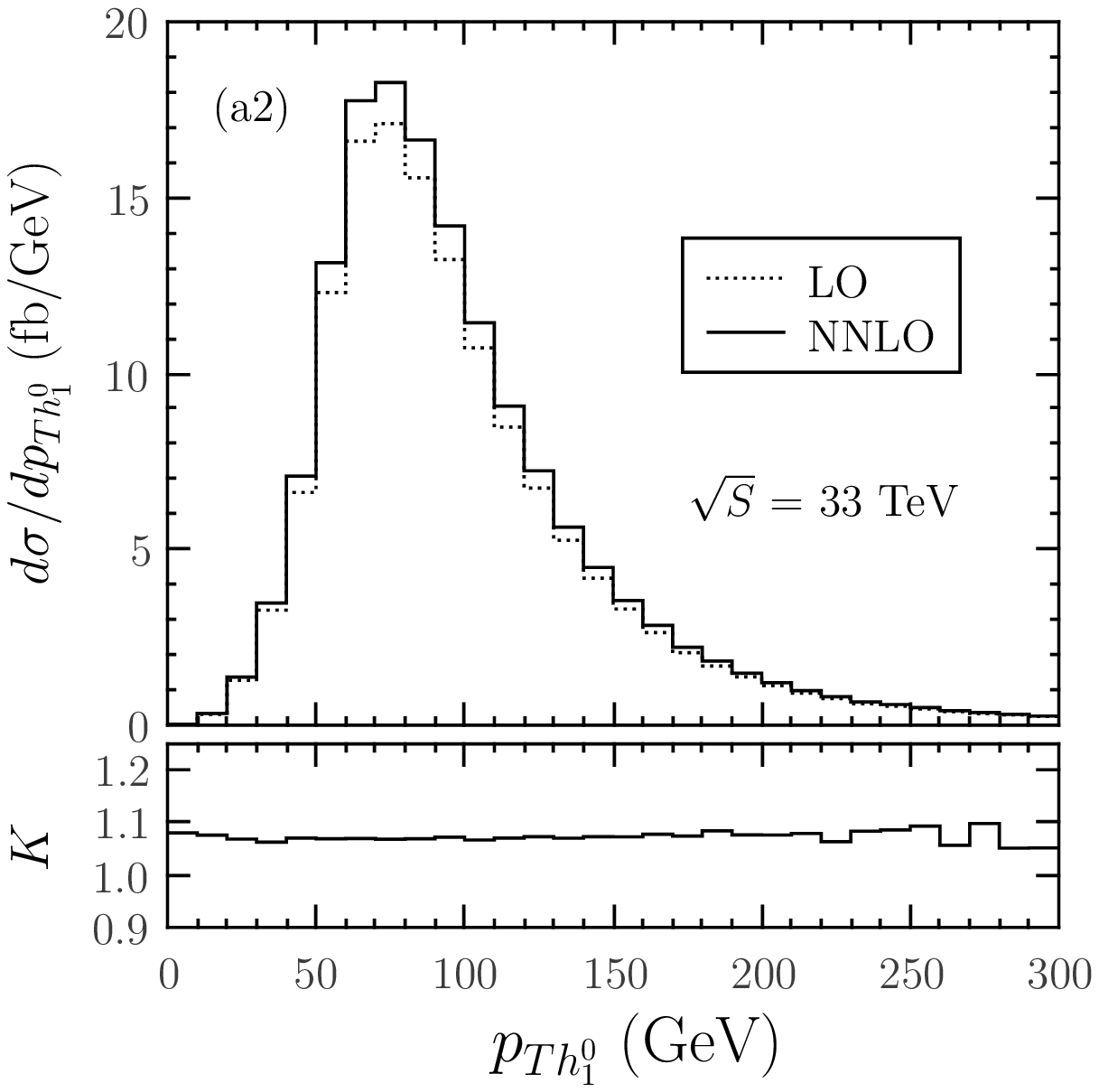}
\includegraphics[scale=0.50]{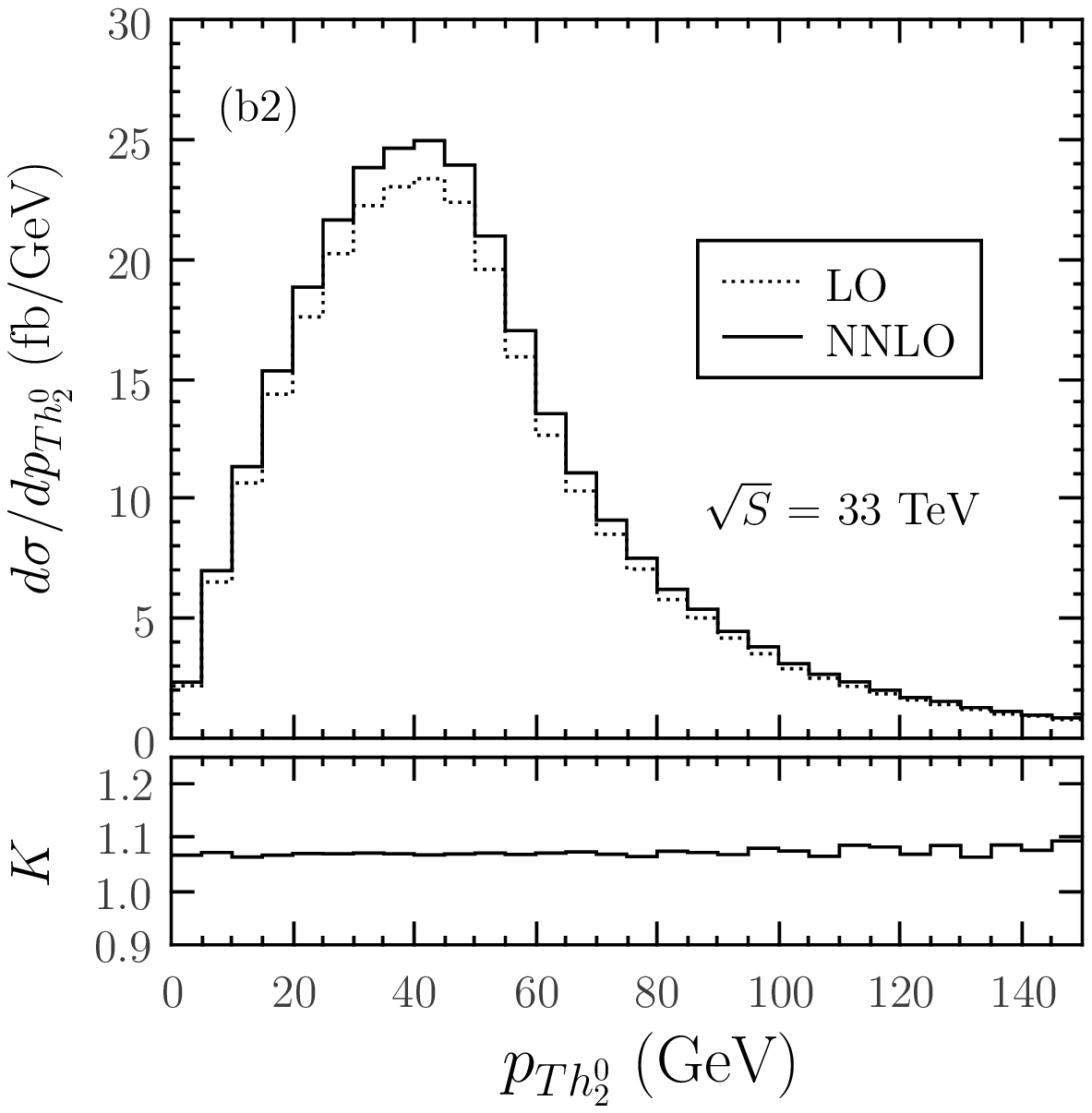}
\includegraphics[scale=0.50]{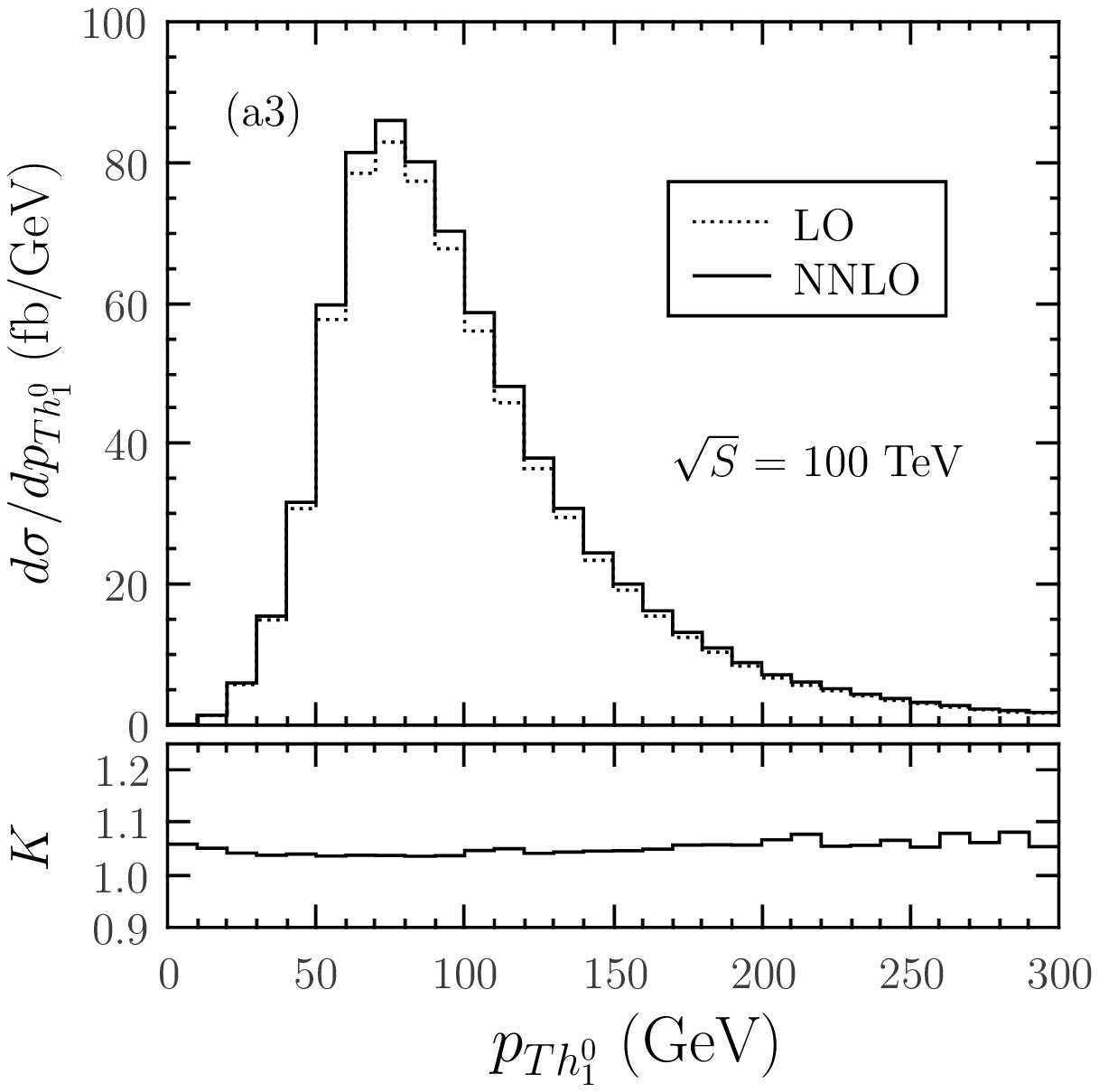}
\includegraphics[scale=0.50]{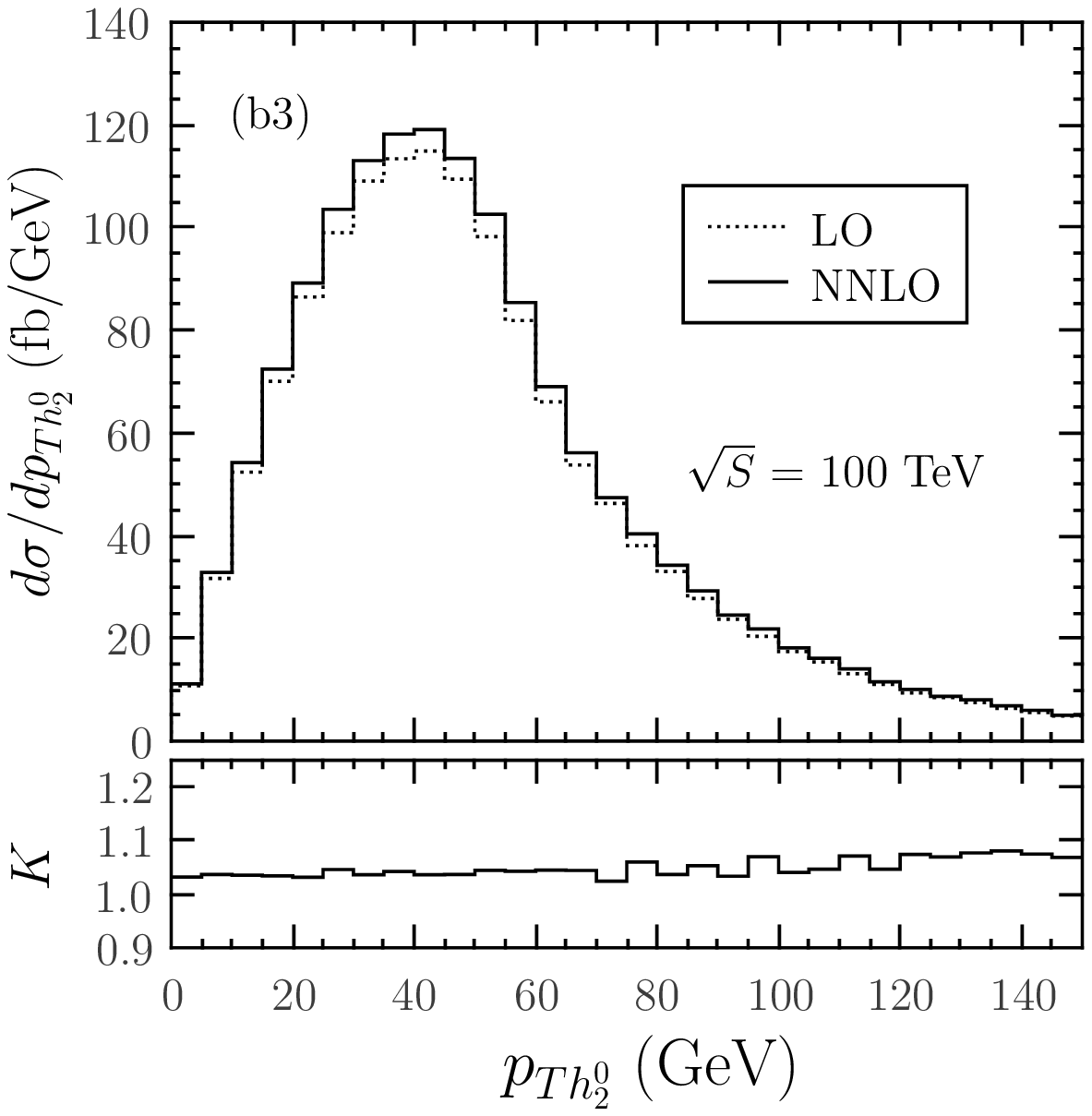}
\caption{The LO, NNLO QCD corrected Higgs transverse momentum
distributions and corresponding $K$-factors for the VBF $h^0 h^0 +
2~ jets$ production at $\sqrt{S} = 14$, $33$ and $100~ {\rm TeV}$
$pp$ colliders at benchmark point B1. (a1), (a2) and (a3) are for
the leading Higgs boson. (b1), (b2) and (b3) are for the second
Higgs boson.} \label{fig:Higgs-pt-distribution}
\end{figure}

The LO and NNLO QCD corrected transverse momentum distributions of
the leading Higgs boson $h^0_1$ and the second Higgs boson $h^0_2$
at $\sqrt{S} = 14$, $33$ and $100~ {\rm TeV}$ $pp$ colliders at
benchmark point B1 are shown in
Figs.\ref{fig:Higgs-pt-distribution}(a1, a2, a3) and (b1, b2, b3),
respectively, where the leading Higgs boson $h^0_1$ and the second
Higgs boson $h^0_2$ are defined as
\begin{eqnarray}
p_{Th^0_{1}} > p_{Th^0_{2}}~.
\end{eqnarray}
We see from these figures that the NNLO QCD corrections can enhance
the Higgs $p_T$ distributions, but the $K$-factors are less than
$1.10$ for both $p_{Th^0_{1}}$ and $p_{Th^0_{2}}$ distributions in
the plotted $p_T$ range. The $p_{Th^0_{1}}$ distributions reach
their maxima at $p_{Th^0_{1}} \sim 80~{\rm GeV}$, while
the $p_{Th^0_{2}}$ distributions reach their maxima at $p_{Th^0_{2}}
\sim 45~{\rm GeV}$, at $\sqrt{S} = 14,~ 33,~ 100~ {\rm
TeV}$ $pp$ colliders, respectively.

\par
\begin{figure}[ht!]
\centering
\includegraphics[scale=0.50]{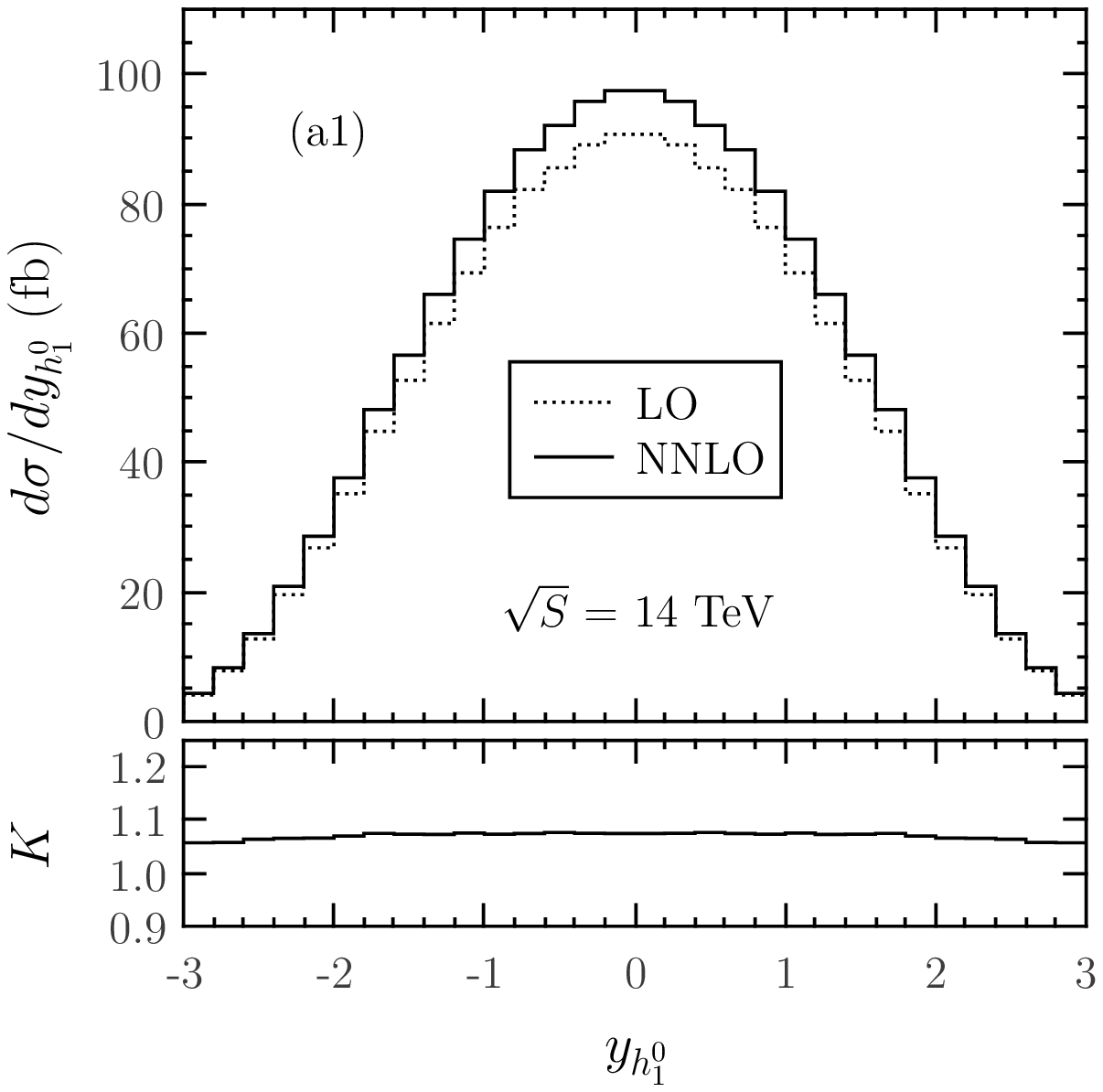}
\includegraphics[scale=0.50]{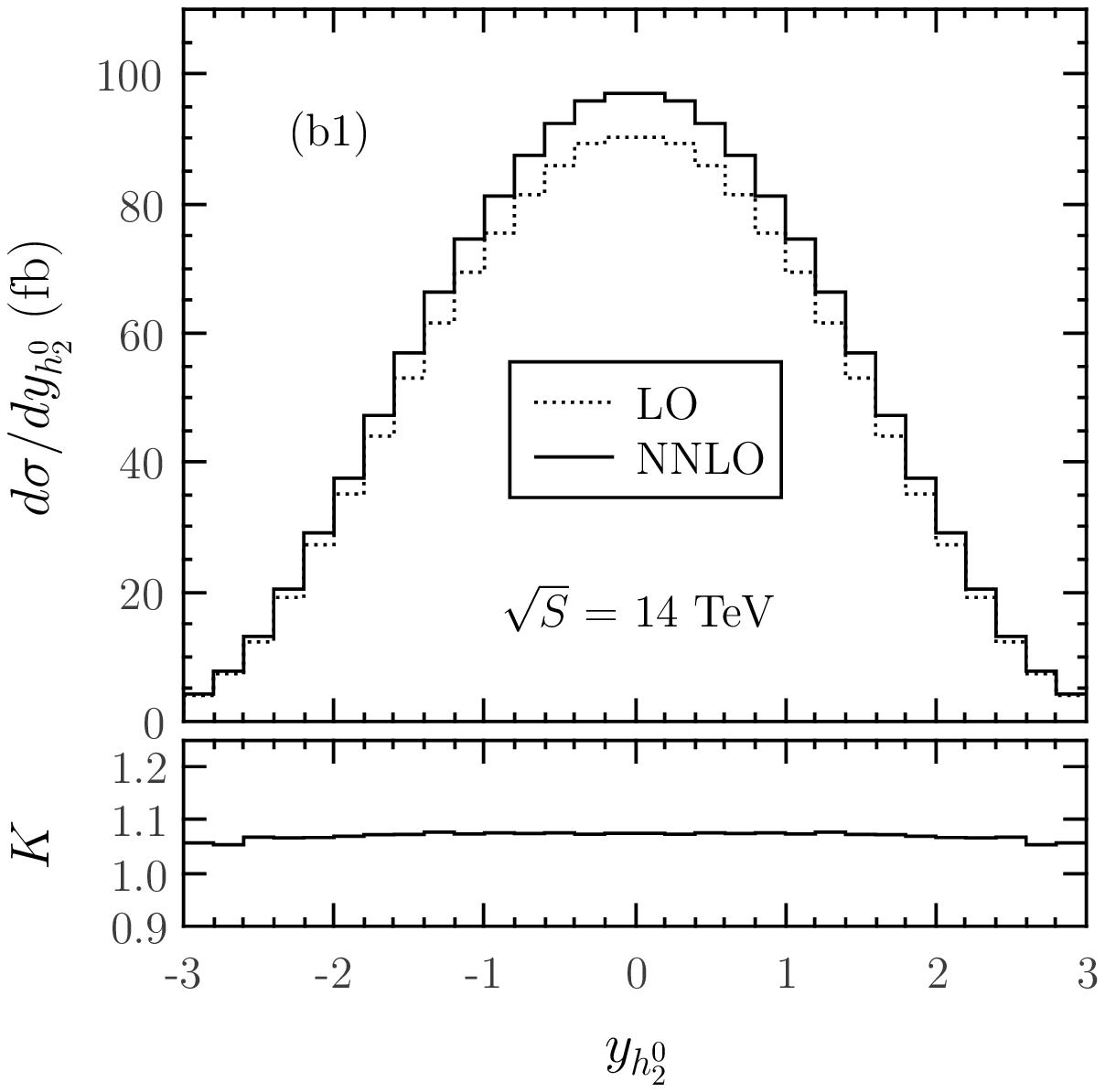}
\includegraphics[scale=0.50]{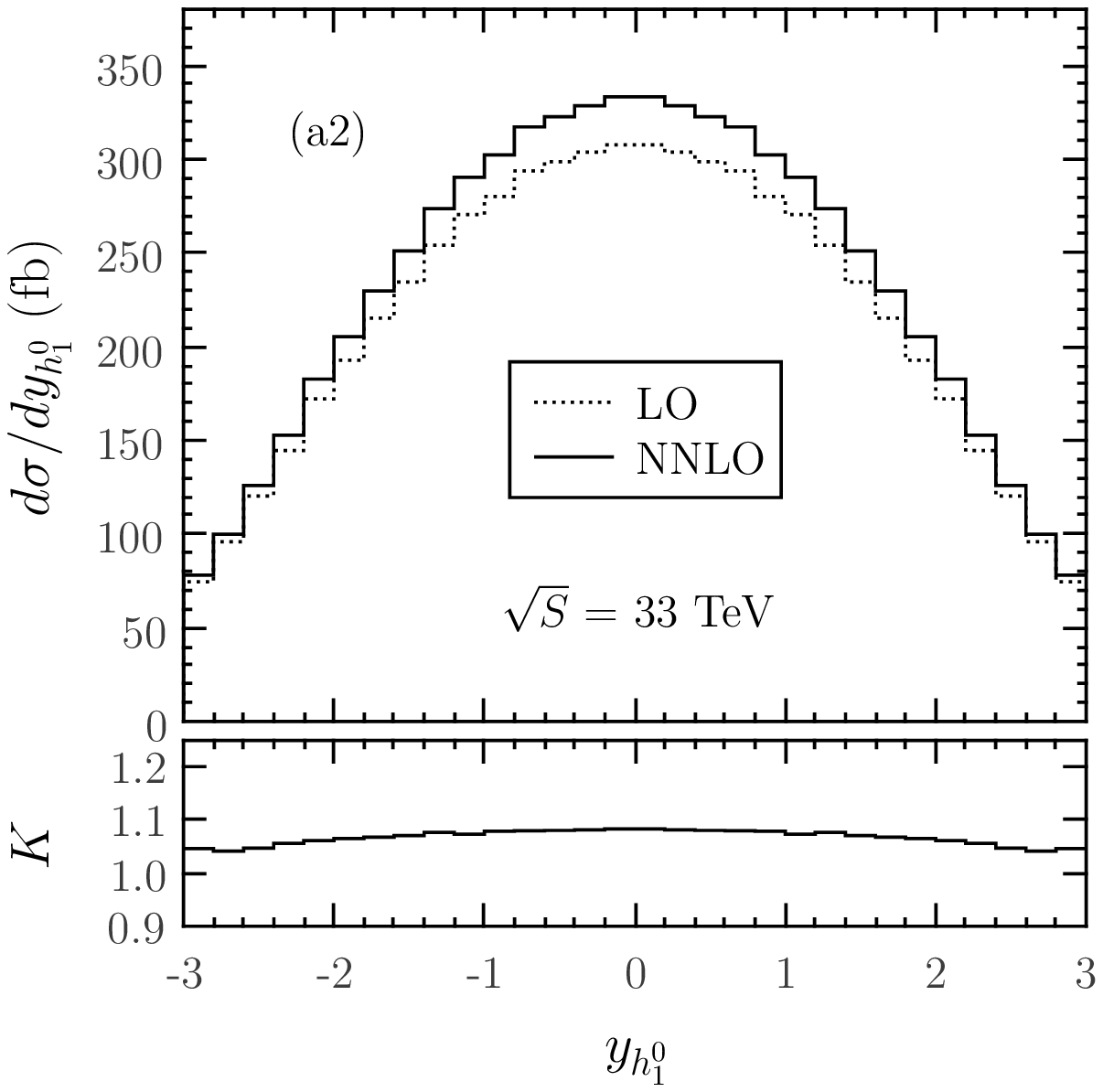}
\includegraphics[scale=0.50]{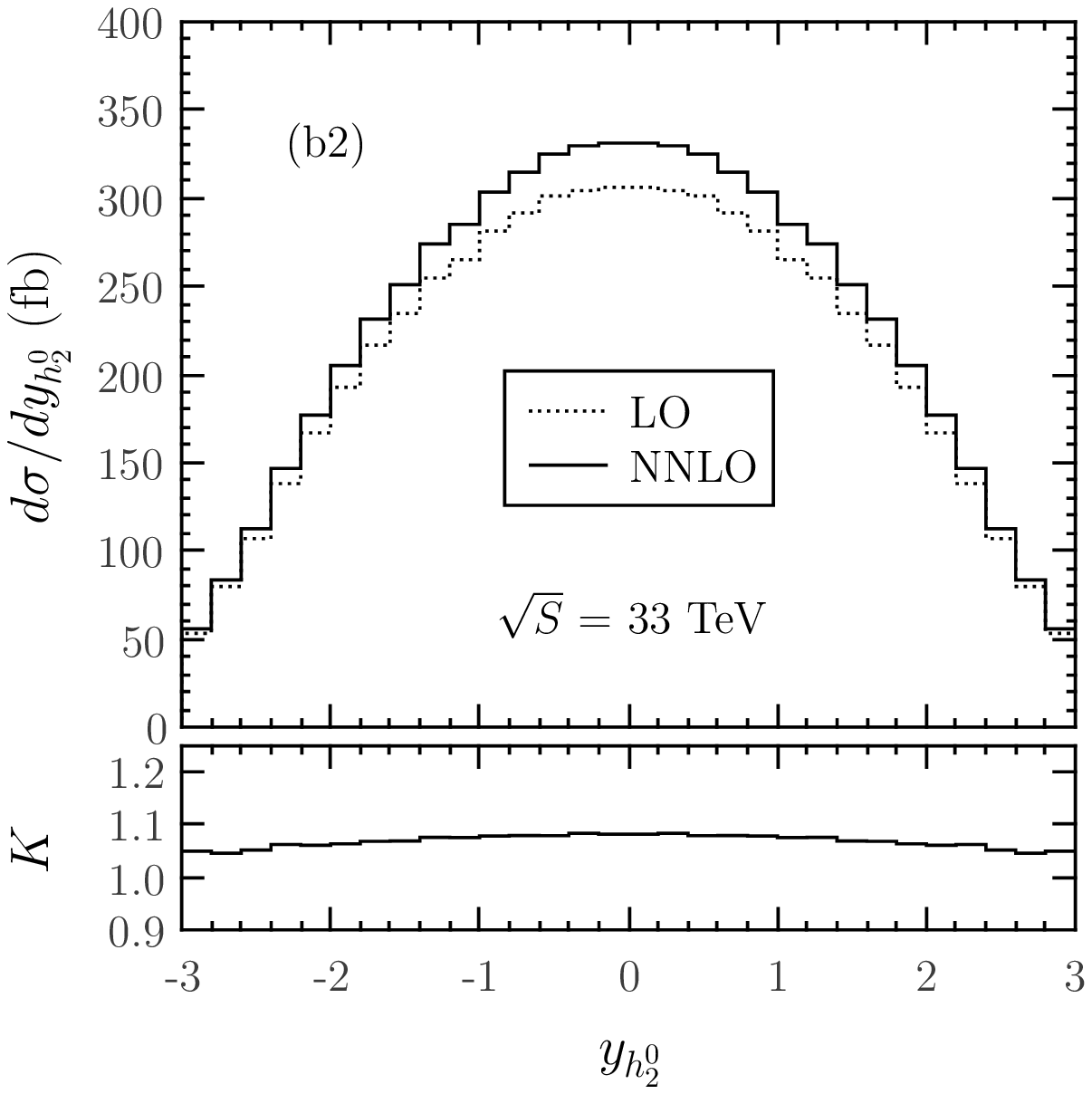}
\includegraphics[scale=0.50]{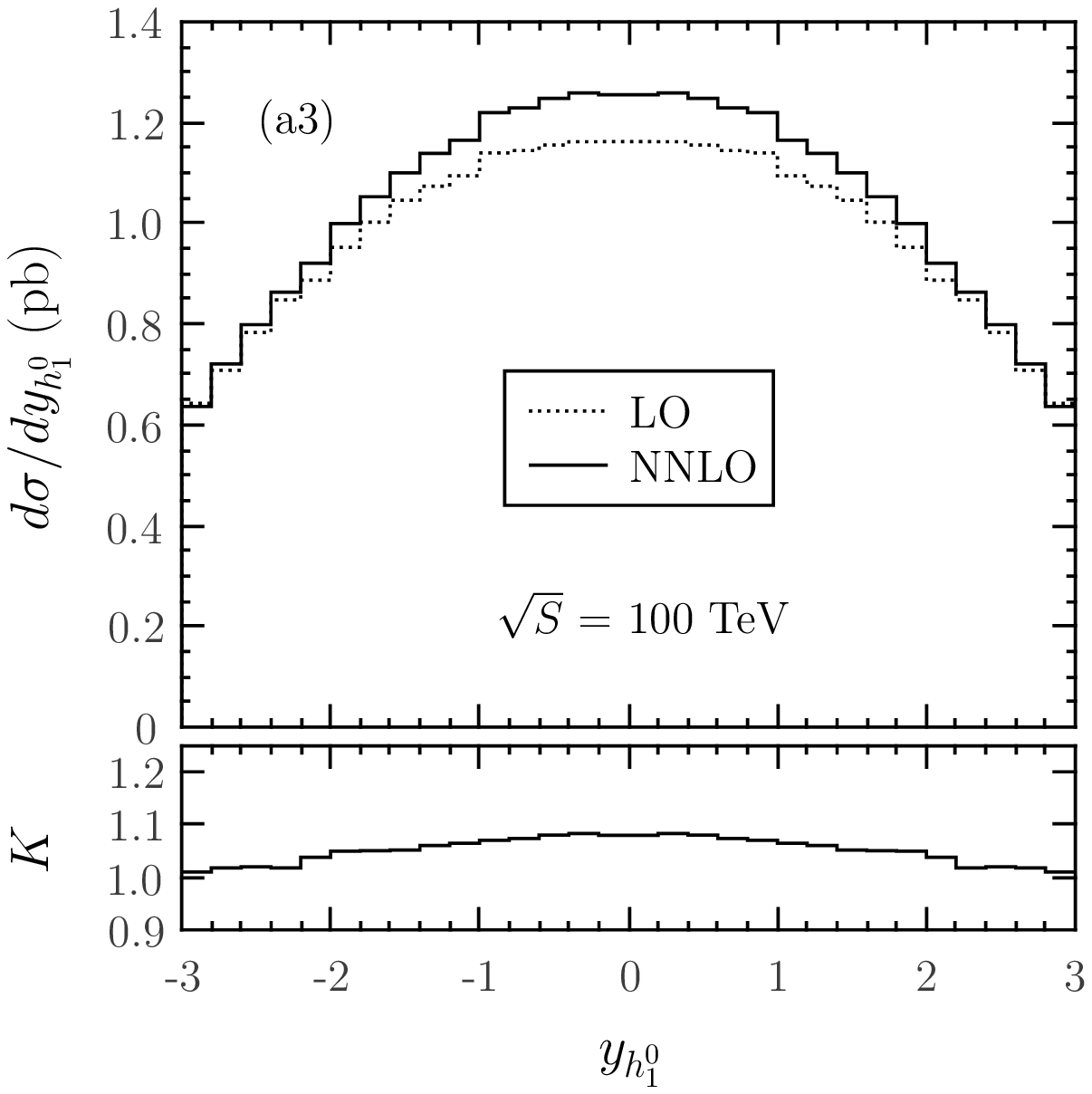}
\includegraphics[scale=0.50]{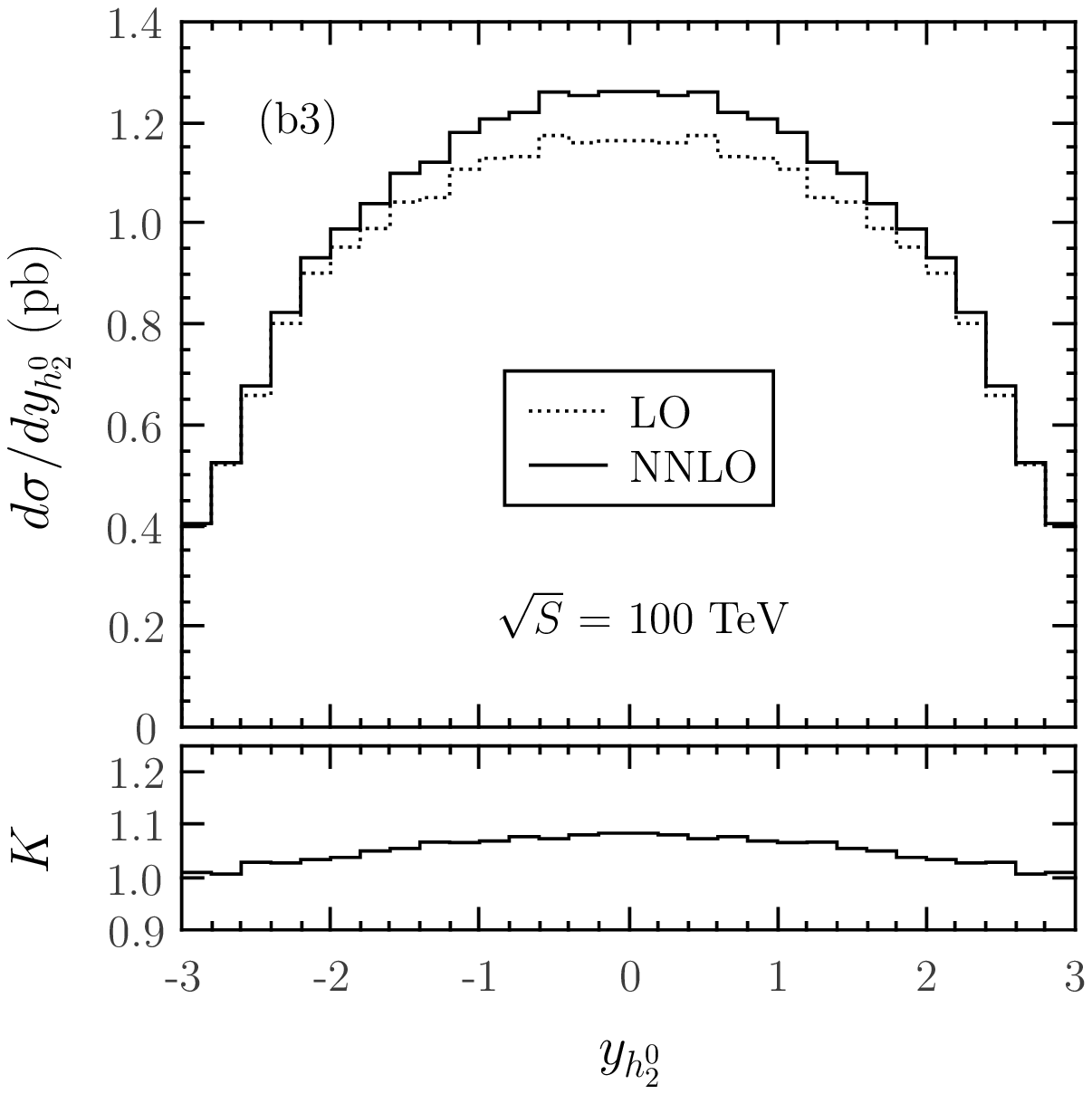}
\caption{The LO, NNLO QCD corrected Higgs rapidity distributions and
corresponding $K$-factors for VBF $h^0 h^0 + 2~ jets$ production at
$\sqrt{S} = 14$, $33$ and $100~ {\rm TeV}$ $pp$ colliders at
benchmark point B1. (a1), (a2) and (a3) are for the leading Higgs
boson. (b1), (b2) and (b3) are for the second Higgs boson.}
\label{fig:Higgs-rapidity-distribution}
\end{figure}

The LO and NNLO QCD corrected rapidity distributions of the leading
Higgs $h^0_1$ and the second Higgs $h^0_2$ at $\sqrt{S} = 14$, $33$
and $100~ {\rm TeV}$ $pp$ colliders at benchmark point B1 are
plotted in Figs.\ref{fig:Higgs-rapidity-distribution}(a1, a2, a3)
and (b1, b2, b3), respectively. From these figures we can see that
the two final Higgs bosons prefer to be produced in the central
rapidity region.

\par
\begin{figure}[ht!]
\centering
\includegraphics[scale=0.50]{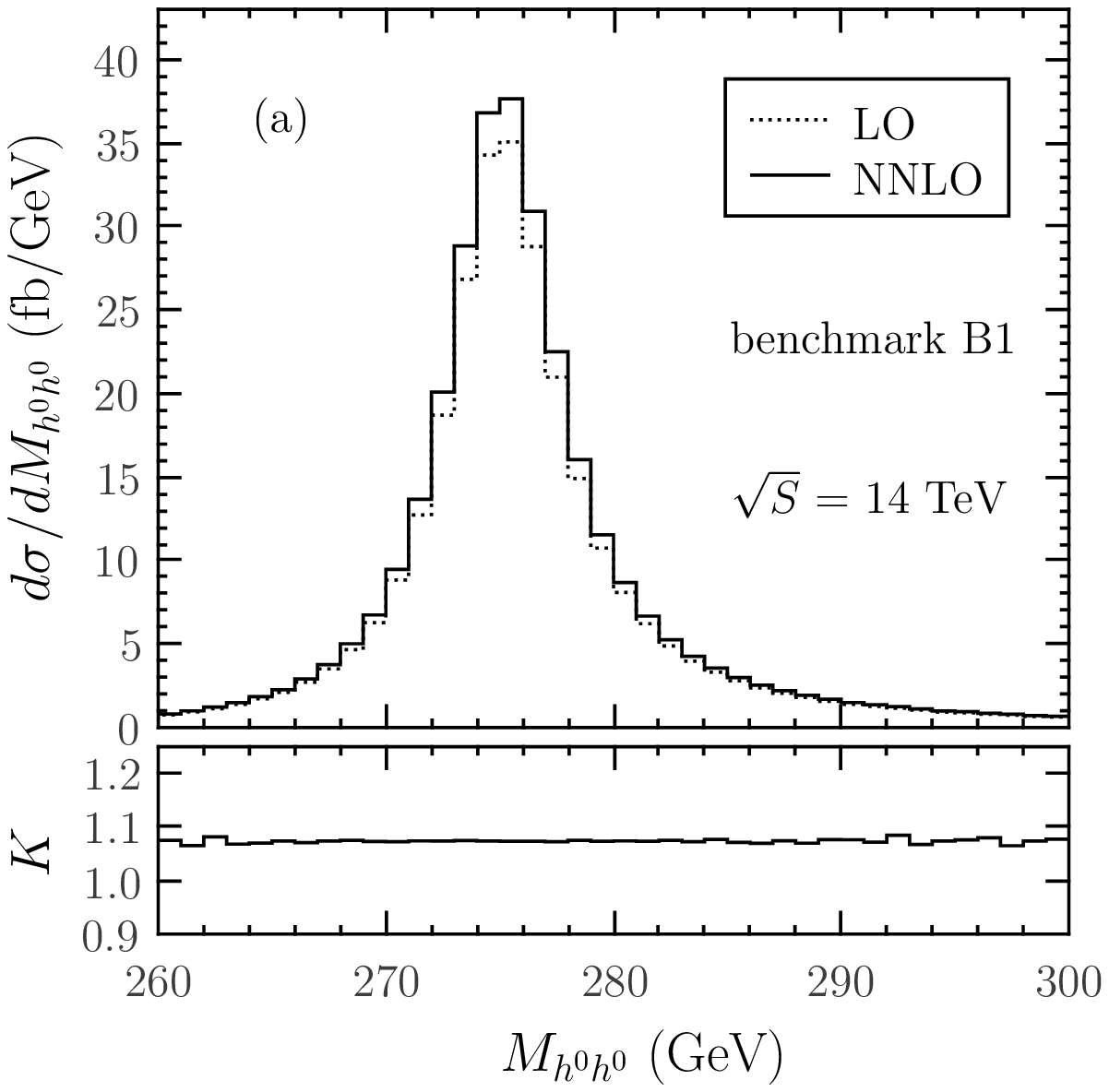}
\includegraphics[scale=0.50]{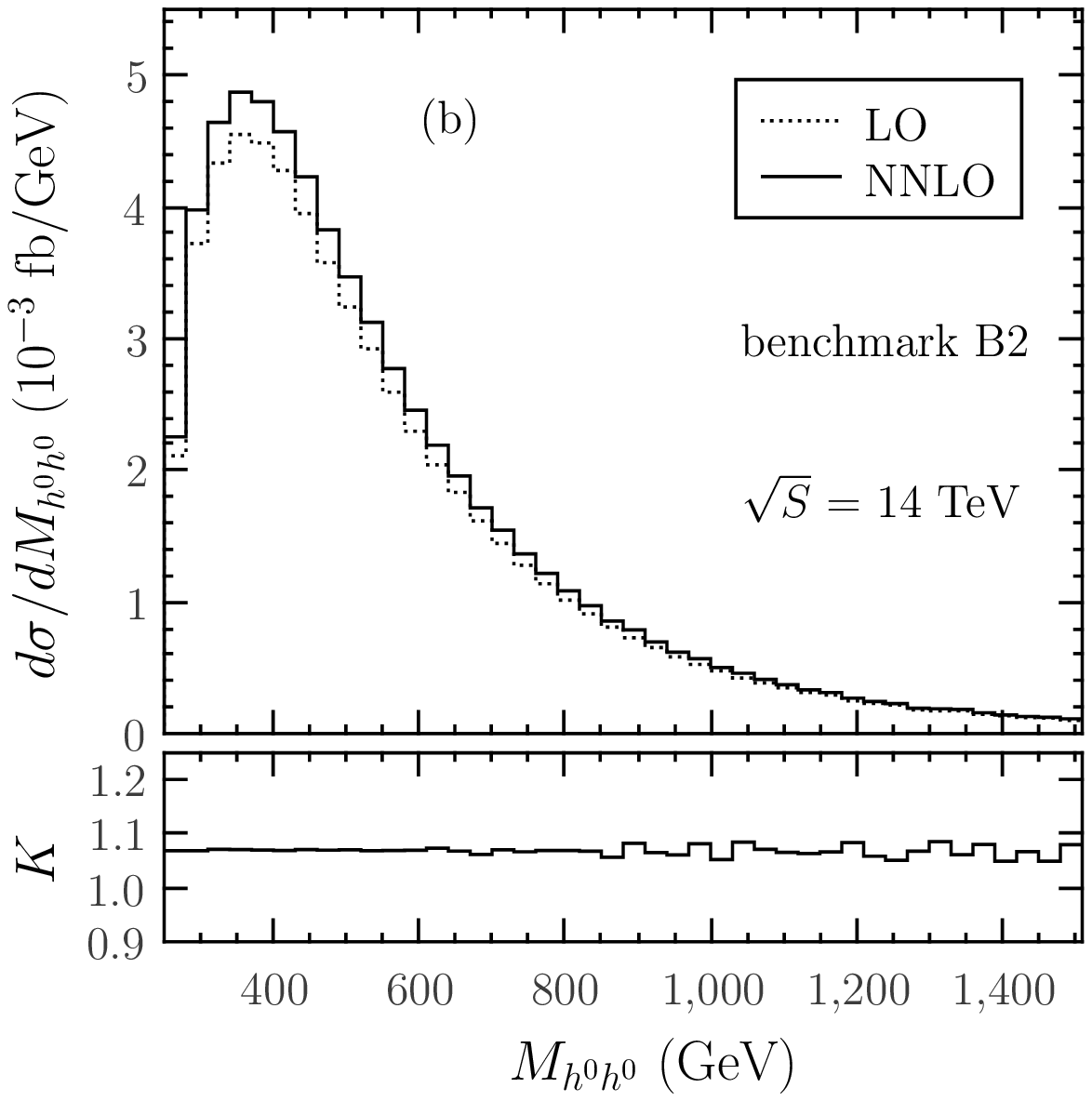}
\caption{The LO, NNLO QCD corrected Higgs pair invariant mass
distributions and corresponding $K$-factors for the VBF $h^0 h^0 +
2~ jets$ production at $14~ {\rm TeV}$ LHC. (a) at the benchmark
point B1. (b) at the benchmark point B2.}
\label{fig:Higgs-invmass-distribution}
\end{figure}

The LO and NNLO QCD corrected distributions of the invariant mass of
final Higgs boson pair, $\frac{d \sigma_{(NN)LO}}{d M_{h^0h^0}}$, at
the $\sqrt{S} = 14~ {\rm TeV}$ LHC at benchmark points B1 and B2 are
given in Figs.\ref{fig:Higgs-invmass-distribution}(a) and (b),
respectively. For benchmark B2, the Higgs pair invariant mass
distributions are mostly concentrated in the vicinity of $M_{h^0h^0}
\sim 370~ {\rm GeV}$, and then decrease slowly with the increment of
$M_{h^0h^0}$. The corresponding $K$-factors are in the range of
$[1.05,~ 1.09]$. For benchmark B1, the $M_{h^0h^0}$ distributions
are sharply enhanced at $M_{h^0h^0} \sim 276~ {\rm GeV}$ due to the
$H^0$ resonance effect. The total cross section for the VBF $h^0h^0
+ 2~ jets$ production is dominated by the VBF $H^0 + 2~ jets$
production mechanism with subsequent decay of $H^0 \to h^0 h^0$. By
analyzing the invariant mass distribution of $h^0$-pair, we can
directly probe the $\lambda_{h^0h^0h^0}$ and $\lambda_{H^0h^0h^0}$
trilinear Higgs self-couplings which can reconstruct the Higgs
potential, and extract the resonance $H^0$ production signal.

\vskip 5mm
\section{Summary}
\label{sec:summary}
\par
In this paper, we investigated in detail the light, $CP$-even Higgs
pair production via VBF at $pp$ colliders within the type-II 2HDM up
to the QCD NNLO by adopting the structure function approach. We
studied the dependence of the integrated cross section on the model
parameters. To assess the theoretical uncertainty on the
perturbative predictions, we considered both the
factorization/renormalization scale uncertainty and the combined
PDF+$\alpha_s$ uncertainty. Our numerical results show that the
scale uncertainty is comparable with the combined PDF+$\alpha_s$
uncertainty, and can be reduced significantly by the QCD NNLO
corrections. At the $\sqrt{S}=14~{\rm TeV}$ LHC the total QCD NNLO
corrected theoretical upper and lower deviations, defined as the
linear combination of the scale and combined PDF+$\alpha_s$
uncertainties, are below $5.1\%$. We study also the kinematic
distributions of the final Higgs bosons at the QCD NNLO by using the
structure function approach, and obtain the phase space dependent
$K$-factor. By analyzing $M_{h^0h^0}$ distribution, we could obtain
the strength of the $\lambda_{H^0h^0h^0}$ coupling relative to the
$\lambda_{h^0h^0h^0}$ coupling qualitatively, and extract the
resonance $H^0$ production signal which provides a means of probing
the extended Higgs sector.

\vskip 5mm
\par
\noindent{\large\bf Acknowledgments:} This work was supported in
part by the National Natural Science Foundation of China (Grants.
No.11275190, No.11375008, No.11375171), and the Fundamental Research
Funds for the Central Universities (Grant. No.WK2030040044).

\vskip 10mm
\appendix
\renewcommand{\theequation}{\arabic{equation}}
\setcounter{equation}{0}

\begin{center}{\bf{\Large  Appendix}}
\end{center}

\renewcommand{\theequation}{A.\arabic{equation}}
\setcounter{equation}{0}

\vskip 5mm
\section{Phase space element for VBF Higgs pair production process }
\label{sec:PhaseSpace}
\par
Here we briefly document the parameterization for the phase space of the VBF Higgs pair production
process. As shown in Fig.\ref{fig:vbf}, We denote the momenta of the incoming protons and outgoing
proton remnants as $P_{i}$ and $P_{X_i}$ $(i = 1, 2)$, respectively. The Lorentz invariant four-body
final state phase space element for the VBF Higgs pair production process (shown in Eq.(\ref{eq:sfapproach}))
can be rewritten as
\begin{eqnarray}
d PS =
\left[
\prod_{i=1,2}
d s_i \frac{d^4 P_{X_i}}{\left( 2 \pi \right)^4} 2 \pi \delta \left( P_{X_i}^2 -s_i \right)
\right]
d PS_2(k_1, k_2)
\left( 2 \pi \right)^4 \delta^4 \left( P_1 + P_2 - P_{X_1} - P_{X_2} - \sum_{j=1,2} k_j \right),~~
\end{eqnarray}
where $k_1$ and $k_2$ are the momenta of the two outgoing Higgs bosons, and
\begin{eqnarray}
d PS_2(k_1, k_2) =
\frac{d^3 \vec{k}_1}{\left( 2 \pi \right)^3 2 E_1}
\frac{d^3 \vec{k}_2}{\left( 2 \pi \right)^3 2 E_2}.
\end{eqnarray}
By integrating out the invariant masses of proton remnants $s_1$ and $s_2$, and replacing the integral variables
$P_{X_i}$ with $q_i = P_{X_i} - P_i$, we obtain
\begin{eqnarray}
d PS =
\left[
\prod_{i=1,2}
\frac{d^4 q_i}{\left( 2 \pi \right)^3}
\right]
d PS_2(k_1, k_2)
\left( 2 \pi \right)^4 \delta^4 \left( q_1 + q_2 + \sum_{j=1,2} k_j \right).
\end{eqnarray}
The integration measures can be expressed as
\begin{eqnarray}
d^4 q_1 = \frac{Q_1^2}{x_1} d x_1 \frac{d^3 \vec{k}_3}{2 E_3},~~~~~~~~~
d^4 q_2 = \frac{Q_2^2}{x_2} d x_2 \frac{d^3 \vec{k}_4}{2 E_4},
\end{eqnarray}
where the light-like momenta $k_3$ and $k_4$ are defined as
\begin{eqnarray}
k_3 = q_1 + x_1 P_1,~~~~~~~~
k_4 = q_2 + x_2 P_2.
\end{eqnarray}
At the end we can express the phase space element in terms of the
DIS variables $x_i~ (i = 1, 2)$ and the three-momenta of final Higgs
bosons and the $\vec{k}_3$ and $\vec{k}_4$ as
\begin{eqnarray}
d PS =
\frac{Q_1^2 Q_2^2}{x_1 x_2} d x_1 d x_2
\left[
\prod_{i=1}^4
\frac{d^3 \vec{k}_i}{\left( 2 \pi \right)^3 2 E_i}
\right]
\left( 2 \pi \right)^4 \delta^4 \left( x_1 P_1 + x_2 P_2 - \sum_{i=1}^4 k_i \right).
\end{eqnarray}

\renewcommand{\theequation}{B.\arabic{equation}}
\setcounter{equation}{0}
\vskip 5mm
\section{Matrix element ${\cal M}^{\mu \nu}$ for $VV \to h^0 h^0$ subprocess}
\label{sec:matrix-element}
The matrix element for the $Z(-q_1) + Z(-q_2) \rightarrow h^0(k_1) + h^0(k_2)$ process can be written as
\begin{eqnarray}
\label{eq:ZZfusion}
{\cal M}^{\mu\nu}_{ZZ}
&=&
i \sqrt{2} G_F M_Z^2
\left\{
g^{\mu\nu}
\left[
2 + 4 \sin^2(\beta - \alpha)
\left(
\frac{M_Z^2}{(q_1 + k_1)^2 - M_Z^2} + \frac{M_Z^2}{(q_1 + k_2)^2 - M_Z^2}
\right)
\right.
\right. \nonumber \\
&&
+
\left.
\frac{6 \sin(\beta - \alpha) \left(2 \cos(\beta + \alpha) + \sin2\alpha \sin(\beta - \alpha)\right)}{\sin 2\beta}
\frac{m_{h^0}^2}{(q_1 + q_2)^2 - m_{h^0}^2}
\right. \nonumber \\
&&
+
\left.
\frac{2 \cos^2(\beta - \alpha) \sin2\alpha}{\sin 2\beta}
\frac{m_{H^0}^2 + 2 m_{h^0}^2}{(q_1 + q_2)^2 - m_{H^0}^2}
\right] \nonumber \\
&&
+
\left.
(q_1 + 2 k_1)^{\mu} (q_2 + 2 k_2)^{\nu}
\left[
\frac{\cos^2(\beta - \alpha)}{(q_1 + k_1)^2-m_{A^0}^2} + \frac{\sin^2(\beta - \alpha)}{(q_1 + k_1)^2 - M_Z^2}
\right]
\right. \nonumber \\
&&
+
\left.
(q_1 + 2 k_2)^{\mu} (q_2 + 2 k_1)^{\nu}
\left[
\frac{\cos^2(\beta - \alpha)}{(q_1 + k_2)^2-m_{A^0}^2} + \frac{\sin^2(\beta - \alpha)}{(q_1 + k_2)^2 - M_Z^2}
\right]
\right\}.
\end{eqnarray}
For the $W^+(-q_1) + W^-(-q_2) \to h^0(k_1) + h^0(k_2)$ process, the
matrix element is expressed as
\begin{eqnarray} \label{eq:WWfusion}
{\cal M}^{\mu\nu}_{W^+W^-} &=& i \sqrt{2} G_F M_W^2 \left\{
g^{\mu\nu} \left[ 2 + 4 \sin^2(\beta - \alpha) \left(
\frac{M_W^2}{(q_1 + k_1)^2 - M_W^2} + \frac{M_W^2}{(q_1 + k_2)^2 -
M_W^2} \right) \right.
\right. \nonumber \\
&&
+
\left.
\frac{6 \sin(\beta - \alpha) \left(2 \cos(\beta + \alpha) + \sin2\alpha \sin(\beta - \alpha)\right)}{\sin 2\beta}
\frac{m_{h^0}^2}{(q_1 + q_2)^2 - m_{h^0}^2}
\right. \nonumber \\
&&
+
\left.
\frac{2 \cos^2(\beta - \alpha) \sin2\alpha}{\sin 2\beta}
\frac{m_{H^0}^2 + 2 m_{h^0}^2}{(q_1 + q_2)^2 - m_{H^0}^2}
\right] \nonumber \\
&&
+
\left.
(q_1 + 2 k_1)^{\mu} (q_2 + 2 k_2)^{\nu}
\left[
\frac{\cos^2(\beta - \alpha)}{(q_1 + k_1)^2-m_{H^{\pm}}^2} + \frac{\sin^2(\beta - \alpha)}{(q_1 + k_1)^2 - M_W^2}
\right]
\right. \nonumber \\
&&
+
\left.
(q_1 + 2 k_2)^{\mu} (q_2 + 2 k_1)^{\nu}
\left[
\frac{\cos^2(\beta - \alpha)}{(q_1 + k_2)^2-m_{H^{\pm}}^2} + \frac{\sin^2(\beta - \alpha)}{(q_1 + k_2)^2 - M_W^2}
\right]
\right\}.
\end{eqnarray}

\par
For the $W^-(-q_1) + W^+(-q_2) \rightarrow h^0(k_1) + h^0(k_2)$
process, we have ${\cal M}^{\mu\nu}_{W^-W^+} = {\cal
M}^{\mu\nu}_{W^+W^-}$. In the region of $m_{H^0} > 2 m_{h^0}$, the
complex pole scheme is applied.

\end{document}